# Measurement of the Fine Structure Constant as a Test of the Standard Model

Richard H Parker[1,*], Chenghui Yu[1,*], Weicheng Zhong[1], Brian Estey[1], and Holger Müller[1,2†]

[1]Department of Physics, 366 Le Conte Hall MS 7300, University of California, Berkeley, California 94720, USA.

[2]Lawrence Berkeley National Laboratory, One Cyclotron Road, Berkeley, CA, 94720, USA.

*These authors contributed equally to this work.

†Correspondence to: hm@berkeley.edu.

**Abstract**: Measurements of the fine structure constant α, using methods from atomic, condensed-matter, and particle physics, are powerful tests of the overall consistency of theory and experiment across physics. We have measured α = 1/137.035999046(27), at $2.0\times10^{-10}$ accuracy, via the recoil frequency of cesium-133 atoms in a matter-wave interferometer. We used multiphoton interactions such as Bragg diffraction and Bloch oscillations to increase the phase difference for the interferometer to over 12 million radians, which reduced the statistical uncertainty and enabled control of systematic effects at the 0.12 part-per-billion level. This is the most accurate measurement of the fine structure constant and demonstrates the largest phase of any Ramsey-Bordé atom interferometer. It is the first time a comparison between the electron g-2 measured with a Penning trap and derived from an atom interferometer via the Standard Model is limited by the uncertainty in the g-2 measurement. The 2.5-sigma tension in the comparison rejects dark photons as the reason for the unexplained anomaly of the muon's magnetic moment at 99% confidence level. Its implications for multiple dark sector candidates as well as substructure of the electron may be a sign of physics beyond the standard model and warrant further investigation.

The fine structure constant α characterizes the strength of the electromagnetic interaction. It has been measured using methods from diverse fields of physics (Fig. 1), whose agreement is a remarkable confirmation of the consistency of theory and experiment across physics. In particular, α can be obtained from measurements of the electron's gyromagnetic anomaly $g_e$-2 by using the standard model of particle physics, including quantum electrodynamics to fifth order (involving >10,000 Feynman diagrams) and muonic as well as hadronic physics (*1, 3*). This path leads to 0.24 part-per-billion (ppb) accuracy (*2, 8, 10*), and was up to now the most accurate measurement of α.

An independent measurement of α at comparable accuracy creates an opportunity to test the standard model. The most accurate of previous such measurements have been based on the kinetic energy $\hbar^2 k^2/(2m_{\mathrm{At}})$ of an atom of mass $m_{\mathrm{At}}$ that recoils from scattering a photon of momentum $\hbar k$ (*3*), where $\hbar$ is the reduced Planck constant and $k=2\pi/\lambda$ is the laser wavenumber (where $\lambda$ is the laser wavelength). Experiments of this type yield $\hbar/m_{\mathrm{At}}$, and have measured α to 0.62 ppb (*4*) via the relation

$$\alpha^2 = \frac{2R_\infty}{c}\frac{m_{\mathrm{At}}}{m_e}\frac{h}{m_{\mathrm{At}}}.$$

The Rydberg constant $R_\infty$ is known to 0.006 ppb accuracy (*10*), and the atom-to-electron mass ratio is known to better than 0.1 ppb for many species.

The fundamental tool of our experiment is a matter-wave interferometer (*11, 12*). Similar to an optical interferometer, it splits waves from a coherent source along different paths, recombines them, and measures the resulting interference to extract the phase difference accumulated between the waves on the paths. Sequences of laser pulses are used to direct and recombine the atomic matter waves along different trajectories, to form a closed interferometer (*13*). The phase evolution is governed by the Compton frequency of the atoms. The probability of detecting each atom at the output of the interferometers is a function of the phase accumulated between the different paths; measuring the total atom population in each output estimates this phase. For the Ramsey-Bordé interferometer geometry used in this experiment, the phase is proportional to the photon recoil energy, and can therefore be used to measure the ratio $\hbar/m_{\mathrm{Cs}}$, and from that the fine structure constant α.

Our experiment uses a number of methods to increase the signal and suppress systematic errors. We use 10-photon processes as beam splitters for the matter waves, which increases the recoil energy by a factor of 25 relative to standard 2-photon Raman processes (*14*). To accelerate the atoms by up to another 800 $\hbar k$ (400 $\hbar k$ up, 400 $\hbar k$ down), we apply a matter-wave accelerator: atoms are loaded into an optical lattice, a standing wave generated by two laser beams, which is accelerated by ramping the frequency of the lasers ("Bloch oscillations") (*4*, *15*). Coriolis-force compensation suppresses the effect of Earth's rotation. In addition, we have applied ac-Stark shift compensation (*16, 17*) and demonstrated a new spatial-filtering technique to reduce sources of decoherence, further enhance the sensitivity, and suppress systematic phase shifts. An end-to-end simulation of the experiment was run (*15*) to help us identify and reduce systematic errors, and help confirm the error budget. To avoid possible bias, we adopted a blind measurement protocol which was un-blinded only at the end. Combining with precise measurements of the cesium (*18*) and electron (*19*) mass, we find

$$\alpha^{-1} = 137.035999046(27),$$

with a statistical uncertainty of 0.16 ppb and a systematic uncertainty of 0.12 ppb (0.20 ppb total). The measurement of $h/m_{\mathrm{Cs}} = 3.0023694721(12) \times 10^{-9}$ m$^2$/s also provides an absolute mass standard in the context of the proposed new definition of the kilogram (*13*). This proposed definition will assign a fixed numerical value to the Planck constant, to which mass measurements could then be linked through measurements of h/M, such as this one, via Avogadro spheres. Our result agrees within 1 sigma with previous recoil measurements (*4*) and has a 2.5 sigma tension with measurements (2*, 8, 10*) based on the gyromagnetic moment.

Our matter-wave interferometer is based on the one described in (*15*), in which cesium atoms are loaded in a magneto-optical trap, launched upward in an atomic fountain, and detected as they fall back down—the interferometer sequence occurs during the parabolic flight. Fig. 2 shows the trajectories of an atom wave packet in our experiment, formed by impulses from pairs of vertical, counter-propagating laser pulses on the atoms. Each pulse transfers the momentum of $2n$=10 photons with 50% probability by multiphoton Bragg diffraction, acting as a beam splitter for matter waves. Bragg diffraction allows for large momentum transfer at each beam splitter, creating a pair of atom wave packets that separate with a velocity of about 35 mm/s. After a time interval $T$, a similar pulse splits the wave packets again, creating one pair that moves upwards and one that moves down.

The third and fourth pulses recombine the respective paths to form two interferometers. In between the second and the third pulse, we accelerate the atom groups further from one another, using Bloch oscillations in accelerated optical lattices, to increases the sensitivity and suppress systematic effects. This transfers $+2N\hbar k$ of momentum to the upper interferometer and $-2N\hbar k$ to the lower (*16*).

The phase difference between the interferometer arms arises as a result of the kinetic energy $(\hbar k)^2/(2m)$ that the atoms gain from the recoil momentum of the photon-atom interactions, and from the phase transferred during the atoms' interaction with the laser beams. Adding the phases of the two interferometers together cancels effects due to gravity and vibrations. To leading order, the overall phase $\Phi$ of the interferometer geometry shown in Fig. 2, is given by (*15, 20*)

$$\Phi = \Delta\phi_1 - \Delta\phi_2 = 16n(n+N)\omega_r T - 2n\omega_m T \qquad (1)$$

where $\Delta\phi_{1,2}$ are the measured phases of the two interferometers individually, $\omega_r = \hbar k^2/(2m)$ is the photon recoil frequency, $T$ is the time between the laser pulses, and $\omega_m$ is the laser frequency difference we choose to apply between the first and second pairs of pulses (Fig. 2). A measurement proceeds by adjusting $\omega_m$ to find the point where $\Phi=0$ so that $\omega_m=8(n+N)\omega_r$. Since the wavenumber $k$ of the laser is related to the laser frequency, this yields $h/m$ and thus $\alpha$. In our measurement, $n = 5$, $N = 125$-$200$ and $T = 5$-$80$ ms, so that $\Phi$ is $10^6$-$10^7$ rad and $\omega_m$ is 2-3 MHz.

Our error budget (Tab. 1) includes the systematic effects considered in the previous rubidium $h/m$ measurement (*4*). These systematic effects are dominant, and a number of methods is used to reduce them (*21*). Our laser frequency is monitored using a frequency comb generator. Effects caused by the finite radius of the laser beam are controlled by a retro-reflection geometry; delivering all components of the beam via the same single-mode optical fiber, an apodizing filter to improve the Gaussian beam shape, selecting only atoms that stay close to the beam axis, and correcting for drift of the beam alignment in real time to further suppress such effects. The gravity gradient has been measured in situ for subtraction by configuring the atom interferometer as a gravity gradiometer (*22, 23, 24*). Keeping atoms in the same internal state while in all interferometer arms reduces the influence of the Zeeman effect to the one of an acceleration gradient, taken out by the gravity gradient measurement. The index of refraction and atom-atom interactions are reduced by the low density of our atomic sample (*21*).

New systematic effects arise from Bragg diffraction, but can be suppressed to levels much smaller than the well-known systematics just mentioned. The potentially largest systematic is the diffraction phase $\Phi_0$, which we have studied in previous work (*15, 16*). It is caused primarily by off-resonant Bragg scattering in the third and fourth laser pulse, where multiple frequencies for the Bragg beams are used to simultaneously address both interferometers (Fig 2). We can therefore suppress it by using a large number $N$ of Bloch oscillations; this increases the velocity of the atoms and thus the Doppler effect, moving the off-resonant component further off-resonance. It also increases the total phase, further reducing the relative size of the systematic. The diffraction phase is nearly independent of the pulse separation time $T$, so we alternate between two or more (usually six) pulse separation times and extrapolate $T \to \infty$.

To determine the residual $T$-dependent diffraction phase, we employ a Monte Carlo simulation and numerically propagate atoms through the interferometer (*16, 21*). We run the experiment at several different pulse separation times, making sure that there is no statistically significant signal for any unaccounted systematic variation. Overall, systematic errors contribute an uncertainty of 0.12 ppb to the measurement of α. Importantly, we correct for systematic effects due to spatial intensity noise (*25*) and deviations of the beam shape from a perfect Gaussian are (*21*).

Figure 3C shows our data, which was collected over the course of 7 months. Each point represents roughly 1 day of data. The signal to noise ratio of our experiment would allow reaching a 0.2-ppb precision in less than 1 day; but extensive data was taken to suppress and control systematic effects. The measurement campaigns were interspersed with additional checks for systematic errors. Each of the datasets includes typically 6 different pulse separation times (9 datasets used 3 $T$'s and 4 used 4 $T$'s), repeated in ~15-minute bins; the fit algorithm allows each bin of data to have a different diffraction phase (as the various experimental parameters may drift slowly over time), but assumes one value of $h/m$ for the entire dataset.

By combining our measurement with theory (*8, 10*) we calculate the standard model prediction for the anomalous magnetic moment of the electron as

$$a(\alpha) = g_e/2\text{-}1 = 0.00115965218161(23).$$

Comparison with the value obtained through direct measurement (*2*) yields a negative $\delta a = a_{\text{meas}} - a(\alpha) = -0.88(0.36) \times 10^{-12}$.

Comparison of our result to previous measurements of α (Fig. 1) has an error bar below the magnitude of the 5th order QED calculations used in the extraction of α from the electron $g_e$-2 measurement, and thus allows us to confront these calculations with experiment.

In addition, our measurement can be used to probe a possible substructure within the electron. An electron whose constituents have mass $m^* \gg m_e$ would result in a modification of the electron magnetic momentum by $\delta a \sim m_e/m^*$. In a chirally-invariant model, the modification scales as $\delta a \sim (m_e/m^*)^2$. Following the treatment in (*26*), the comparison $|\delta a|$ of this measurement of α with the electron $g_e$-2 result places a limit to a substructure at a scale of $m^* > 411{,}000$ TeV/$c^2$ for the simple model and $m^* > 460$ GeV/$c^2$ for the chirally-invariant model (improvements over the previous limits of $m^* > 240{,}000$ TeV/$c^2$ and $m^* > 350$ GeV/$c^2$, respectively).

Precision measurements of α like ours can also help searching for new dark-sector (or hidden-sector) particles (*21*). A hypothetical dark photon *V'*, which couples to the Standard Model as $-\epsilon F_{\mu\nu} V'^{\mu\nu}/2$ and is parameterized by a mixing strength ϵ and a nonzero mass $m_V$, for example, would lead to a nonzero $\delta a$ that is a function of ϵ and $m_V$ (*27*). We can test the existence of dark photons by comparing our data with the electron $g_e$-2 measurement (*2*). The blue area in Fig. 4A shows the parameter space that is inconsistent with our data. We note that dark photons cause a $\delta a > 0$, opposite to the sign measured in both our experiment and the rubidium measurement (*4*). With the improved error of our measurement, this tension has grown. A model consisting of the standard model and dark photons of any mass $m_V$ or ϵ is now incompatible with the data at as high as 99% CL. Constraints on the theory obtained in this fashion (Fig. 4A) include regions not previously bounded by accelerator experiments and do not depend on the assumed decay branching ratios of the dark photon.

A dark axial vector boson characterized by an axial-vector coupling $c_A$ and mass $m_A$, on the other hand, is favored by the data because it would lead to a negative $\delta a$, but we emphasize that the 2.5 σ tension in the data is insufficient to conclude the existence of a new particle (Fig. 4B); the discrepancy between the two methods of measuring α could be a hint of possible physics beyond the standard model that warrants further investigation. The calculated $\delta a$ places limits on the axial vector parameter space from two sides. The allowed region is partially ruled out by other experiments. However, we note that the region of parameter space consistent with our result and anomalous pion decay is also consistent with current accelerator limits, and thus the remaining region of parameter space warrants further study (*27*).

In particular, dark photons are one proposed explanation for the 3.4 σ discrepancy in the muon $g_\mu$-2 with respect to the standard model prediction (*29*). As shown in Fig. 4, we rule out this explanation for nearly all values of $m_V$ and ϵ, rejecting dark photons as an explanation for the discrepancy at the 99% confidence level for any dark photon mass. Comparing precision measurements of α and $g_e$-2 is a broad probe for new physics, and can search for (or exclude) a plethora of other new particles that have been proposed, such as *B*-*L* vector bosons, axial vector-coupled bosons, and scalar and pseudoscalar bosons including those that mix with the Higgs field, such as the relaxion.

**Acknowledgments:** We acknowledge helpful discussions with Osip Schwartz, Paul Hamilton, Maxim Pospelov, Steven Chu, Barry Taylor, Pei Chen Kuan, Shau Yu Lan, Tim Tait, and Eric Copenhaver. We wish to thank Rana Adhikari for being the 'keeper' of the random number used in the blind analysis. We are particularly grateful to Saïda Guellati-Khélifa and Pierre Cladé for bringing to our attention the change in effective Gouy phase due to intensity variations on the beam. **Funding:** This work was supported by the Alfred P. Sloan Foundation (BR-5044), the David and Lucile Packard Foundation (2009-34712), Jet Propulsion Laboratory grant number 1458850, 1483242, 1531033, and 1553641, National Institute of Standards & Technology award No. 60NANB9D9169, NSF CAREER award No. PHY-1056620, MRI award No. PHY-0923445, and University of California Office of the President award No. 040219. **Author Contributions:** All authors contributed jointly to all aspects of this work. **Competing Interests:** The authors declare that they have no competing interests. **Data and Materials Availability:** All data needed to evaluate the conclusions in the paper are present in the paper and/or the Supplementary Materials.

**Supplementary Materials**

Supplementary Text

Figs. S1 to S10

Table S1

Equations S1 to S2

Reference (22)

**Table 1.** Error Budget. For each systematic effect, more discussion can be found in the listed section of the Supplemental Materials.

| Effect | Sect. | $\delta\alpha/\alpha$ (ppb) |
|---|---|---|
| *Laser Frequency* | 1 | -0.24 ± 0.03 |
| *Acceleration Gradient* | 4A | -1.79 ± 0.02 |
| *Gouy phase* | 3 | -2.60 ± 0.03 |
| *Beam Alignment* | 5 | 0.05 ± 0.03 |
| *BO Light Shift* | 6 | 0 ± 0.002 |
| *Density Shift* | 7 | 0 ± 0.003 |
| *Index of Refraction* | 8 | 0 ± 0.03 |
| *Speckle Phase Shift* | 4B | 0 ± 0.04 |
| *Sagnac Effect* | 9 | 0 ± 0.001 |
| *Mod. Frequency Wavenumber* | 10 | 0 ± 0.001 |
| *Thermal Motion of Atoms* | 11 | 0 ± 0.08 |
| *Non-Gaussian Waveform* | 13 | 0 ± 0.03 |
| *Parasitic Interferometers* | 14 | 0 ± 0.03 |
| *Systematic Error* | | -4.58 ± 0.12 |
| *Statistical Error* | | ± 0.16 |
| *Electron Mass (19)* | | ± 0.02 |
| *Cesium Mass (10,18)* | | ± 0.03 |
| *Rydberg Constant (10)* | | ± 0.003 |
| *Total Uncertainty in α* | | ± 0.20 |

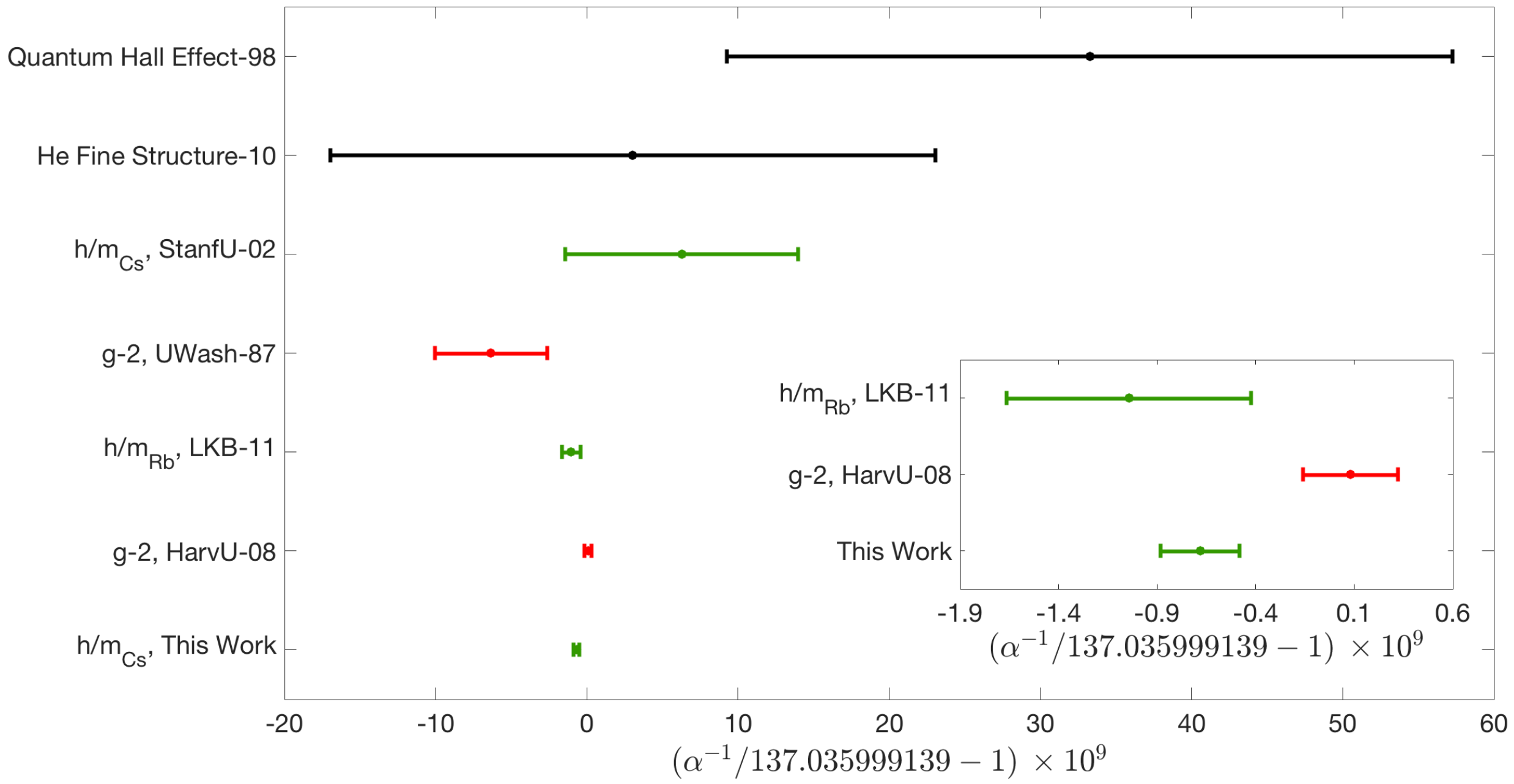

**Fig. 1. Precision measurements of the fine structure constant.** A comparison of measurements (*1, 2, 3, 4, 5, 6, 7, 8*). 'Zero' on the plot is the CODATA 2014 recommended value (*4*). The green points are from photon recoil experiments; the red ones are from electron $g_e$-2 measurements.

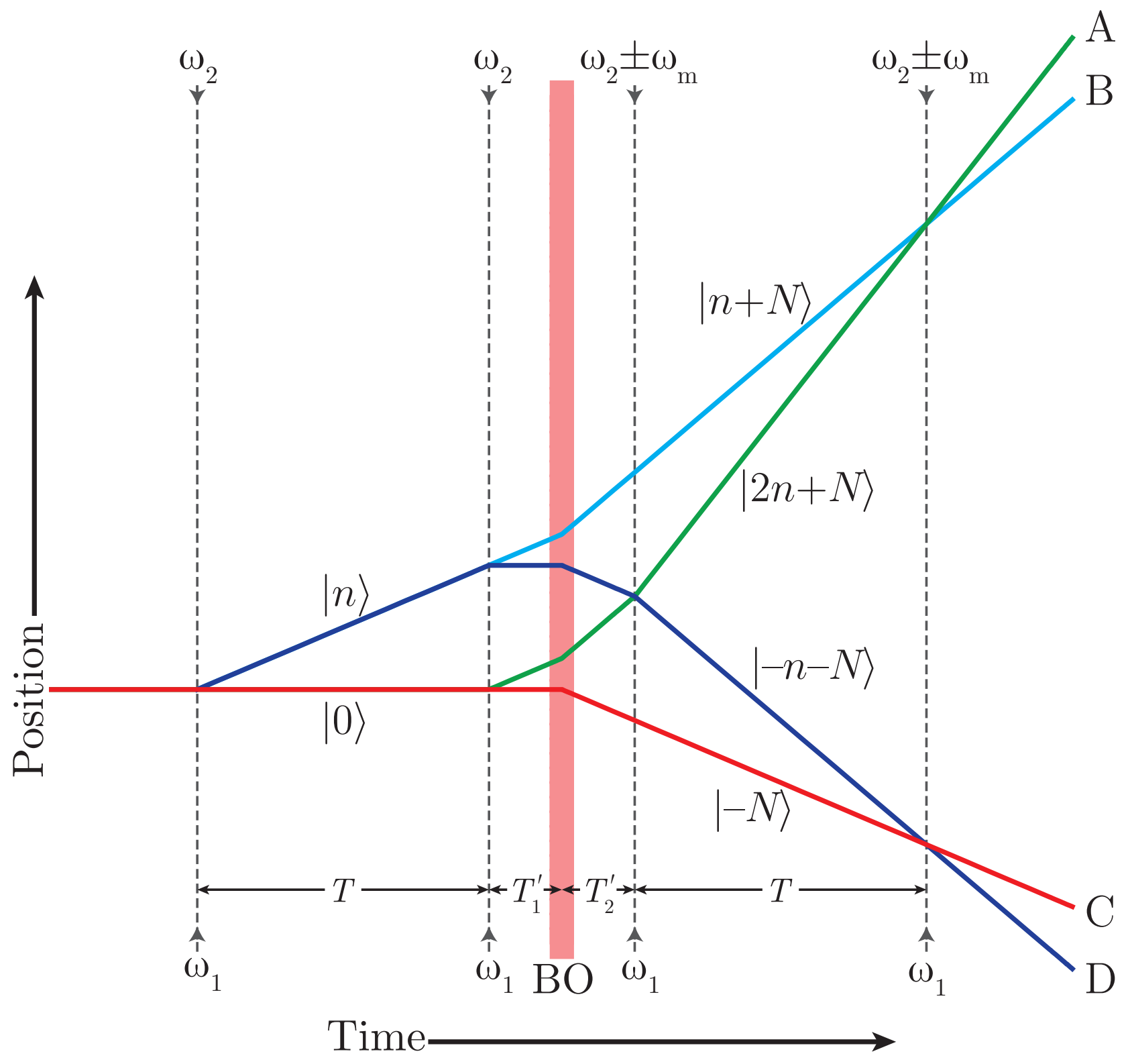

**Fig. 2**. **Simultaneous-Conjugate Atom Interferometers**. The solid lines denote the atoms' trajectories, dashed lines indicate laser pulses with their frequencies indicated. $|n\rangle$ denotes a momentum eigenstate with momentum $2n\hbar k$, where $k$ is the laser wave number. BO; Bloch oscillations. In this figure gravity is neglected.

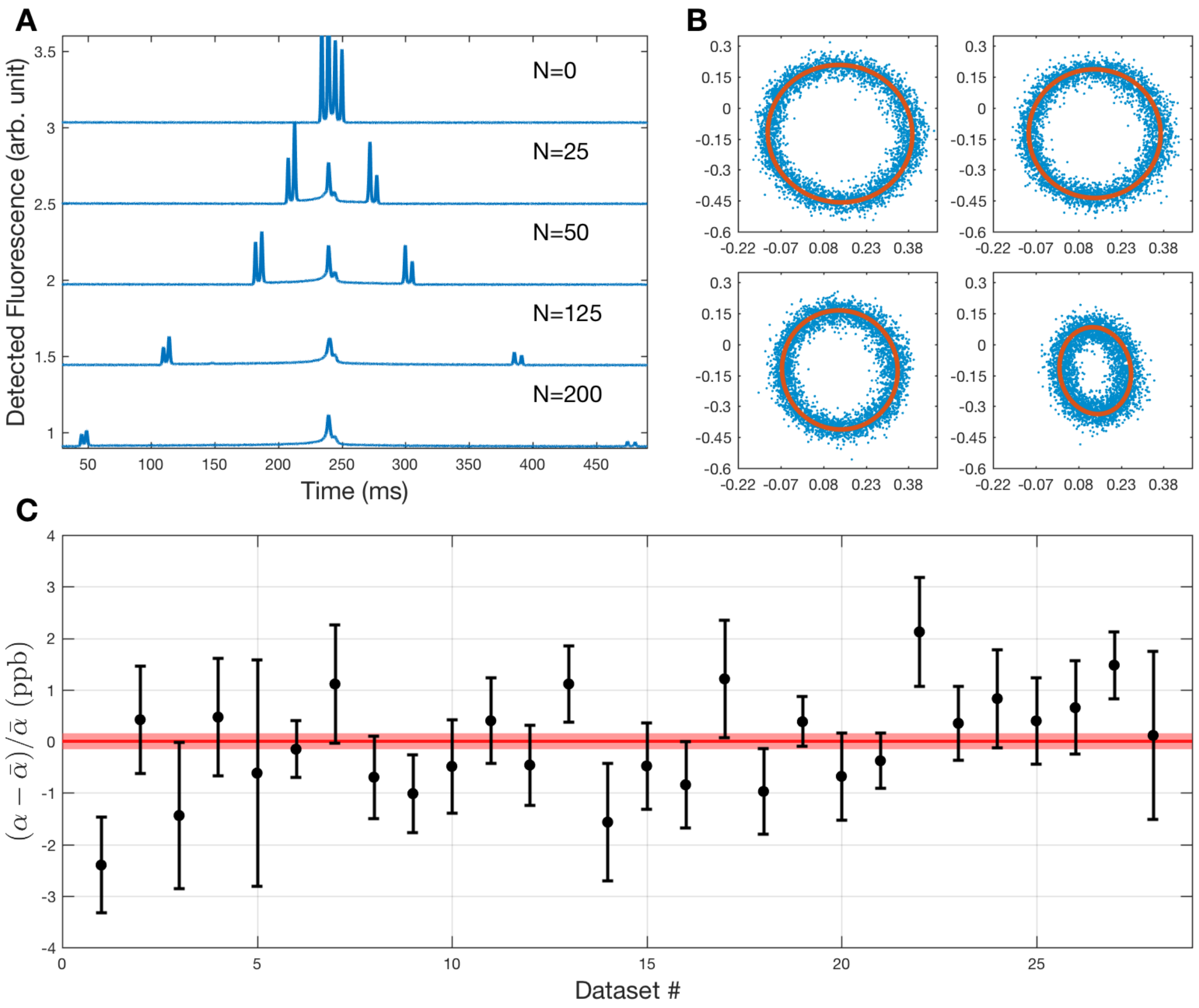

**Fig. 3. Data Analysis. A)** Fluorescence signals of the atom clouds as they fall through the detection region, after the interferometer sequence, for varying number $N$ of Bloch oscillations, measured with fixed laser power and acceleration of the atoms during Bloch oscillations. For visibility, a vertical offset has been applied to each trace. The four outer peaks correspond to the four outputs A-D (Fig. 2) of the interferometers. Atoms left behind by the Bloch oscillations form the central peaks; they do not contribute to the measurement. $T = 5$ ms for these datasets. **B)** The outputs of each interferometer are normalized and plotted parametrically: the $x$-axis is $(C-D)/(C+D)$ and the $y$-axis is $(A-B)/(A+B)$ (A-D are defined in Figure 2). This produces an ellipse, which is fitted to extract the differential phase. The ellipses shown are for $n$=5, $N$=125 and $T$=5, 20, 40, and 80 ms (for a total interferometer phase of over 10 Mrad), respectively. **C)** The datasets used in the determination of α. The pink band represents the overall ±1 sigma statistical error. The reduced chi-squared for the combined data is 1.2, with a p-value of 0.2.

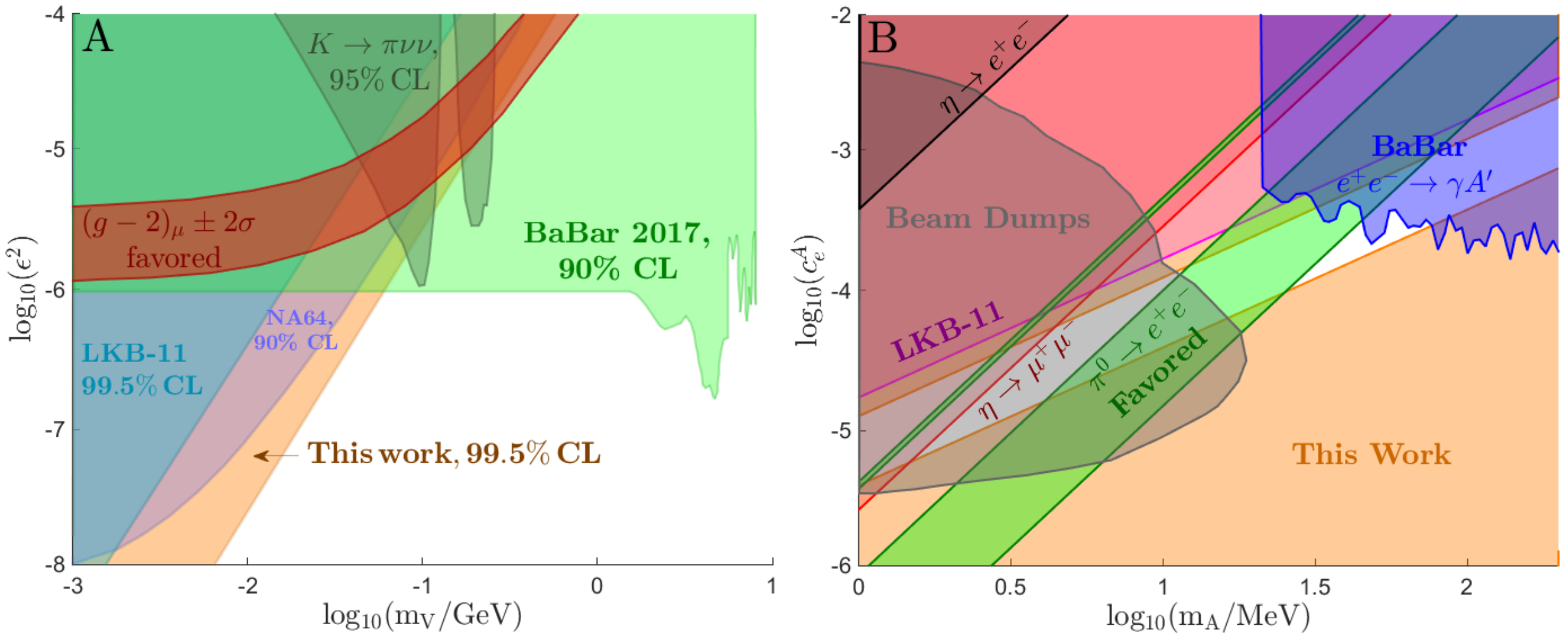

**Fig. 4. Limits on Dark Bosons. A:** Excluded parameter space for dark photons (vector bosons), as a function of the dark photon mass $m_V$ and coupling suppressed by the factor $\epsilon$. The shaded orange and blue regions are ruled out at the indicated CL by comparing the measured $a_e$ (*2, 8, 10*) with that predicted by our $\alpha$ measurement and the LKB11 result (the significance levels have been calculated for a one-tailed test) respectively. The red band is a 95% CL in which the muon $g$-2 is explained by a dark photon. Because our measured $\delta a$ is negative, our measurement disfavors dark photons. Accelerator limits are adapted from Ref 28. **B:** Excluded parameter space for dark axial vector bosons as a function of mass $m_A$ and axial-vector coupling constant $c_A$, whose existence would produce a negative $\delta a$ and is thus favored. Note that our work results in a two-sided bound. The region suggested by anomalous pion decay is shown in green (*27*), at 95% CL. Accelerator limits are adapted from Ref 28.

## Supplementary Text

This PDF file includes:

Supplementary Text

Figs. S1 to S10

Table S1

Equations S1 to S2

Reference (22)

### Section 1: Overview of the Atom Interferometer

Our matter-wave interferometer is similar to the one described in (*15*), see Fig. S1. The atoms are captured from background vapor in a magneto-optical trap, launched using moving optical molasses, and further cooled to about 0.4 μK by polarization gradient cooling and Raman sideband cooling. We select a vertical-velocity subgroup of atoms in the $F$=3, $m_F$=0 sublevel of the 6 $^2S_{1/2}$ electronic ground state (which has no linear Zeeman effect) and discard others.

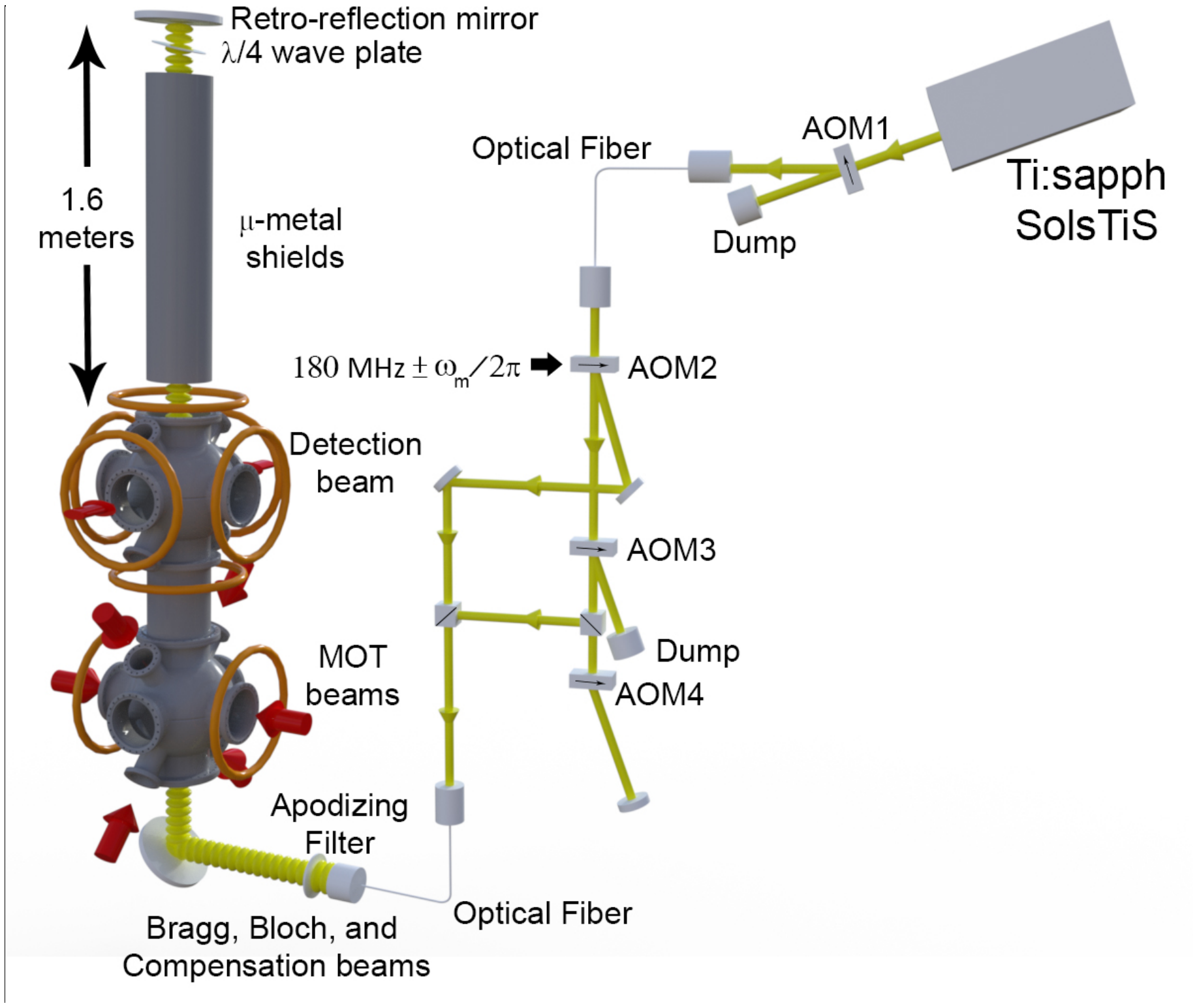

**Figure S1. Diagram of the apparatus.**

All frequencies are referenced to a rubidium frequency standard which is itself stabilized to the global positioning system. A reference laser (New Focus Vortex) is frequency stabilized ("locked") to the 6 $^2S_{1/2}$ $F$=3→ 6 $^2P_{3/2}$ $F'$=2 $D_2$ transition using a hybrid Doppler-free frequency modulation and modulation transfer spectroscopy in a cesium vapor cell. Its frequency is monitored using a femtosecond optical frequency comb (Menlo systems). Variations in the laser power used for the reference spectroscopy will result in correlated variations in the lock frequency. We therefore monitor this laser power and apply a dynamical correction to the laser frequency. The long-term stability of this approach has been found to be better than 10 kHz, which results in a 0.03 ppb uncertainty in α.

To generate the beams for Bragg diffraction and Bloch oscillations, a Coherent 899 titanium:sapphire laser (not shown) is stabilized to the reference laser with a blue offset of about 14 GHz and injection-locks a special MSquared SolsTis titanium:sapphire laser, which is pumped by a Verdi V-18 and achieves over 6 W of output power. An acousto-optic modulator, AOM 1, shapes Gaussian pulses in a closed feedback loop. The power from the laser is then split into two beams by AOM 2. The first beam contains two frequencies, $\omega_\pm = \omega_2 \pm \omega_m$, which are generated by driving AOM 2 with two frequencies. The frequency $\omega_m$ is low enough so that the diffraction efficiencies for the two frequencies are nearly equal. This way of generating the frequency pair ensures that both components have the same optical paths, which results in a constant intensity balance and low differential phase noise between the two components. The second beam is the undiffracted order of AOM 2. It is stripped of any amplitude modulation by AOM 3 and is frequency-shifted by the double pass AOM 4 to a frequency $\omega_1$. To make sure that the atoms see constant laser frequencies in their rest frame as they fall under gravity, we ramp AOM 4 at a rate of about 23 MHz/s and ramp the lock point of the Coherent 899 laser so that the sum frequency $\omega_1+\omega_2$ remains constant.

These two beams are overlapped with orthogonal polarizations in a single-mode optical fiber, and after passing through a quarter-wave plate enter the vacuum chamber. Inside, they are retroreflected with two passes through a quarter-wave plate. Bragg diffractions are driven by the $\{\omega_1^\uparrow, \omega_+^\downarrow\}$ and the $\{\omega_1^\uparrow, \omega_-^\downarrow\}$ frequency pairs, where the arrows denote the direction of propagation of each component in the fountain.

Loss of contrast with increasing interferometer time and Bloch order used to limit us to $N$~75 Bloch oscillations at $T$=80 ms pulse separation time (*17*). To mitigate that, we use an ac-Stark compensation beam, which is overlapped with the main beam in the same single-mode fiber, and has the opposite detuning (*18*). This results in useable contrast at an increased Bloch order of $N$=125 at $T$=80 ms or $N$=200 at $T$=60 ms, see Fig. S2.

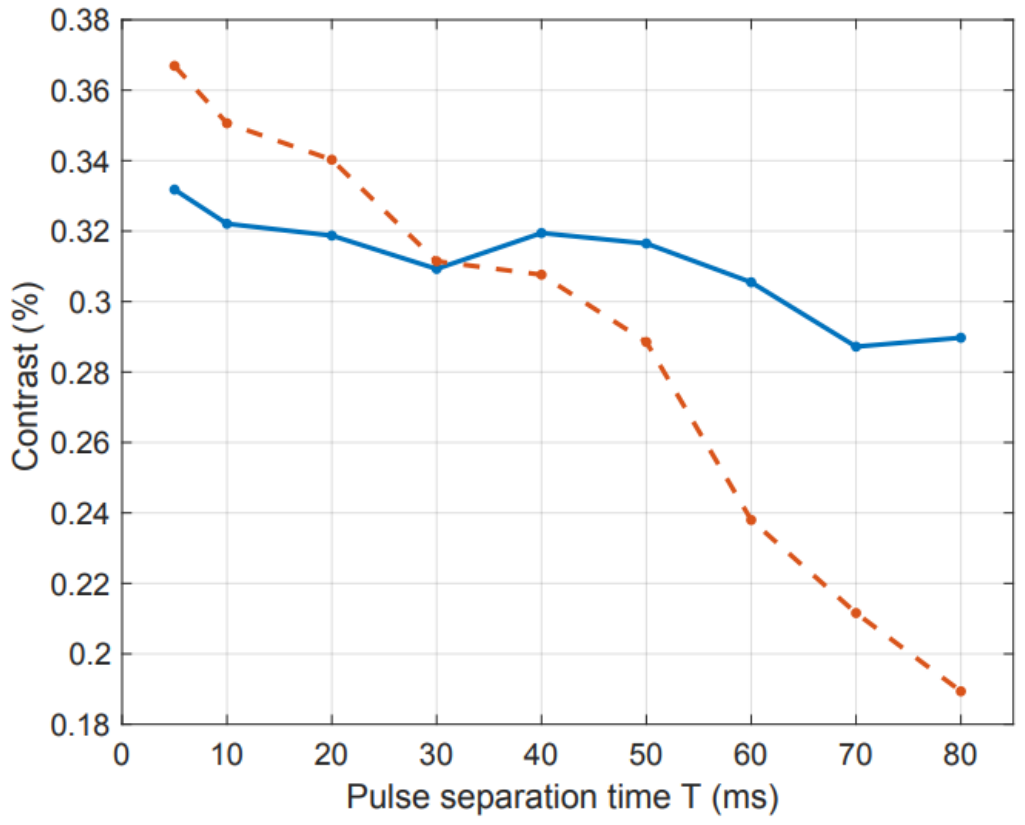

**Figure S2. Contrast vs T.** Contrast in the interferometer as a function of the pulse separation time $T$ is increased, after the addition of the apodizing filter, for $N$=125. Data with (blue, solid line) and without (red, dashed line) ac-Stark compensation is shown. The integration rate of the experiment is proportional to both the contrast and the pulse separation time.

We detect the atoms with a pair of large 100 mm diameter, $f$ = 100 mm lenses, which collect the atom fluorescence and focus the light onto a 2 mm × 2 mm silicon photodetector. Atoms are driven on the cycling 6 $^2S_{1/2}$ $F$=4, $m_F$=4 ↔ 6 $^2P_{3/2}$ $F'$=5, $m_{F'}$=5 transition with circularly polarized light. The fluorescence beam and the detector are apertured so as to detect only the central part of the atomic cloud. A real-time fountain monitor imaging system allows us to know when the fountain drifts out of alignment by imaging the position of the atoms before and after the launch.

The measurement procedure is as follows: an offset is applied to $\omega_m$ so that the overall phase Φ is approximately 90 degrees, so that the ellipse is close to a circle (Fig. 3B). We alternate the offset between ±90 degrees, to cancel any possible systematic effects from ellipse fitting. The Monte Carlo simulation described in Section 11 includes our ellipse fitting algorithm, and therefore any systematics discussed in that section will include any residual systematic effects from ellipse fitting.

One experimental sequence takes 2.4 s, and results in a single fluorescence trace (Fig. 3A). This process is repeated 30 times each with +90 degrees and -90 degrees offset for a pair of ellipses, whose phases are interpolated to extract the $\omega_m$ which would result in an overall zero phase.

This sequence is then repeated for a total of 6 pulse separation times $T$, giving a total of 12 ellipses and six $\omega_m$'s. This is one 'scan' in $T$, which takes ~15 minutes. We continue to scan in $T$ for an entire day, obtaining ~100 scans. The variance of the $\omega_m$'s at each $T$ is used to determine the error. We then fit the set of ~600 $\omega_m$'s with a Levenberg-Marquardt algorithm. Other systematic phase terms, discussed below, are included in the fit function, but have no fit parameters. The overall phase equation used for fitting can be written as

$$\Phi_{\text{Total}} = 16n(n+N)\omega_r T - 2n\tilde{\omega}_m T + \Phi_0 + \Phi_{\text{gradient}} + \Phi_{\delta k} \qquad \text{(S1)}$$

where $\Phi_0$ is the diffraction phase, $\Phi_{\text{gradient}}$ is due to the acceleration gradient and is described in Section 4A, and $\Phi_{\delta k}$ is the phase due to the frequency modulation of the Bragg beams and is described in Section 10. Depending on the frame in which the calculation is done, $\tilde{\omega}_m$ may differ from the $\omega_m$ measured in the lab; see Section 15 for details. The algorithm yields one value of $h/m_{\text{Cs}}$ and its error; this gives one point in Fig. 3C. This was repeated 28 times—the final value of $h/m_{\text{Cs}}$ is obtained by the weighted average of the 28 measurements of $h/m_{\text{Cs}}$ (weighted by the variance). The final statistical uncertainty is calculated by the variance of the mean.

Section 2: Blinding Procedure and avoiding human error

We performed the data-taking and analysis blind, so that our result would not be influenced by knowledge of how ours compared to those of previous measurements. To achieve this, the frequency calibration of the reference laser versus spectroscopy laser power was given to Prof. Rana Adhikari, who added a random offset in the range -1 MHz to +1 MHz, and obfuscated this blinded calibration in a Matlab p-code that prevented the experimenters from deciphering the random offset. This allowed the experimenters to work without knowledge of the exact laser frequency (i.e. 'blind') to within a +/- 3 ppb window.

To reduce the chances of human error, two independent data analysis codes, with different fitting algorithms to extract $h/m$ from each dataset, were used, the only common element being the above obfuscated code. We verified that the two codes gave the same result for the fine structure constant for the same raw data. After all the data was taken and analyzed, Prof. Adhikari provided the random offset to 'unblind' the result, which was then submitted for publication with no further modifications, other than the correction of a typo in the phase calculation that resulted in a 0.2 ppb shift, and the addition of an analysis of the effect of small-scale intensity variations on the Bragg beam (see Section 3).

Section 3: Gouy Phase

The wave-vector of a Gaussian beam and a plane wave differ both on-axis and off-axis. The fractional change in $k_{\text{eff}} = k_1 + k_2$ (where $k_1$ and $k_2$ are the wavenumbers of the up-going and down-going beams respectively) can be expressed as

$$\frac{\delta k_{\text{eff}}}{k_{\text{eff}}} = -\frac{\lambda^2}{2\pi^2 w_0^{\ 2}}\left(1 - \frac{z_0^{\ 2}}{z_R^{\ 2}} - \frac{r^2}{w_0^{\ 2}}\right). \qquad \text{(S2)}$$

The on-axis shift depends on the waist radius $w_0$ of the laser and the position of that waist $z_0$ relative to the retro-reflection mirror (see Fig. S3). The off-axis term depends on the average position $\langle r^2 \rangle$ of the atoms relative to the beam axis and is discussed below.

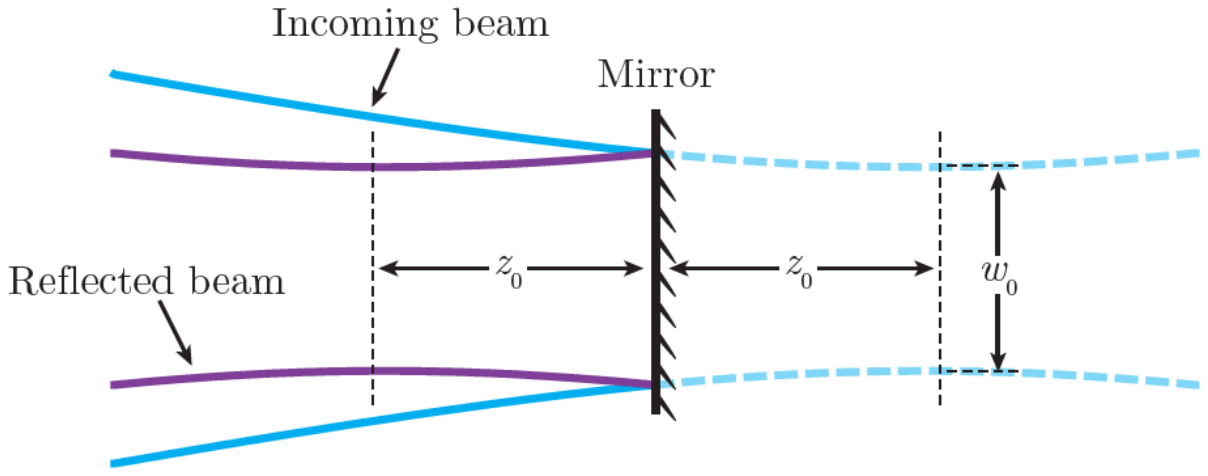

**Fig. S3. Guoy Phase Correction.** A diagram of a Gaussian beam reflecting off the mirror in the chamber. The incoming beam (blue) has a virtual waist at $z = z_0$ and the reflecting beam (purple) has a real waist at $z = -z_0$.

To characterize the Bragg laser beam, we used a CCD camera (Thorlabs BC106N-VIS). In the experiment two different beams were used—a Gaussian beam directly out of an optical fiber with waist 5.7 mm, and an apodized beam designed to remove intensity in the 'wings' (see Section 4B); only data from the latter was used for the final determination of α. We use a 5-pixel-wide mean filter (the pixel pitch is 6.25 μm) to suppress camera pixel noise; the filter is chosen both so that the Monte Carlo contrast matches the experiment, and to agree with the high-spatial-frequency cutoff given by the thermal motion of the atoms during the 10ms Bloch ramp. Measurements of the former beam indicate that a simple Gaussian model adequately describes the profile within one waist of the center, and therefore Eqn. S2 is used to determine the Gouy phase. However, for the apodized the beam is significantly non-Gaussian (Fig. S4).

The change in the effective wavenumber, Eqn S2, can be written as

$$\frac{\delta k}{k}=\frac{1}{2k^2}\frac{\Delta E}{E}-\frac{\langle \rho^2\rangle}{2R^2},$$

where $R$ is the radius of curvature, and $\langle\rho^2\rangle$ is the contrast-weighted mean-square cloud radius. The first term, which is the on-axis Gouy phase contribution, is -2.26 ppb in α. To determine $\langle\rho^2\rangle$, we perform a 3D Monte Carlo simulation (described in more detail in Section 11) which calculates the contribution to the contrast for each atom and calculates the effect of a change to the wavenumber $\delta k_1$, $\delta k_2$, $\delta k_B$, $\delta k_3$, $\delta k_4$ (corresponding to the first, second, Bloch, third, and fourth pulses respectively), resulting in an overall correction to the measured phase in the interferometer given by

$$\delta\Phi=16nN\omega_r T\frac{\delta k_B}{k}+8n^2\omega_r T\frac{\delta k_2}{k}+8n^2\omega_r T\frac{\delta k_3}{k}-16n^2\omega_r T_1'\frac{\delta k_3}{k}-16nN\omega_r T'\frac{\delta k_3}{k}+$$
$$16n^2\omega_r T\frac{\delta k_4}{k}+16nN\omega_r T\frac{\delta k_4}{k}+16n^2\omega_r T_1'\frac{\delta k_4}{k}+16nN\omega_r T_1'\frac{\delta k_4}{k}.$$

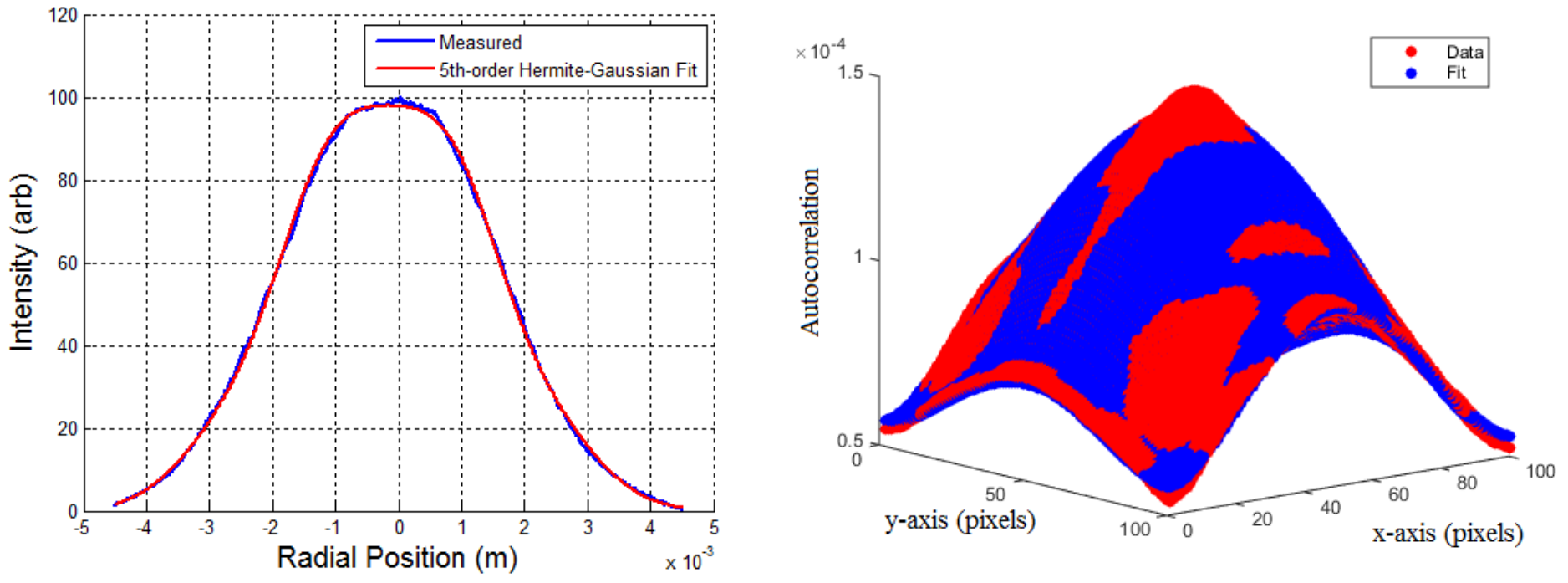

**Fig. S4. Beam profile.** Intensity profile of the apodized beam, measured using the scanning slit beam profiler (left) and the 2-D autocorrelation function of the intensity as determined from a CCD beam profiler. The x and y axes are measured in pixels; 1 pixel corresponds to 6.25μm.

However, it was recently pointed out (*26*) that small-scale intensity variations can also lead to a shift in the effective Gouy phase. Stochastic intensity fluctuations over the cross-section of the laser beam lead to local fluctuation of the photon momentum $p_z$ that are proportional to the two-dimensional Laplacian of the laser beam amplitude $E$,

$$\frac{\delta k}{k} = \frac{1}{2k^2}\frac{\nabla_\perp^2 E}{E}.$$

These fluctuations do not average out completely, since the probability $P(I)$ of an atom to take part in the atom interferometer is a function of the local beam intensity $I$. In particular, the efficiency of Bloch oscillations rises sharply with intensity near a threshold intensity $I_c$. Therefore, we empirically measure Bloch efficiency and include this effect in the Monte Carlo, using CCD images of the Bragg beam which include the small-scale intensity variation. The resulting effective Gouy phase, including both on-axis and off-axis contributions, is -2.60 ± 0.03 ppb in α. The error bar is determined by running the Monte Carlo with several different images, and finding the histogram of the resulting Gouy phases. We have run the Monte Carlo with aggressive smoothing to the images, so that high-frequency intensity noise is not included, and the overall Gouy phase changes by less than 0.1 ppb, providing a bound on the scale of the effect of small-scale intensity variations.

We can verify the scale of the effect of small-scale variations on the beam in two ways: first, with an analytic calculation, and second, by experimental data. For the former, ref (*26*) averages the product $\langle P(I)\nabla^2 E\rangle$ over the beam cross-section, assuming a uniform beam intensity and a Heaviside-shaped $P(I)=\theta(I-I_c)$ that turns on at some critical intensity $I_c$ (*26*). This model, however, is too optimistic for our interferometer, as the influence of intensity noise will rise sharply when the beam intensity $I(r)$ approaches the critical intensity at a certain radius from the center of the beam. We therefore average the effect over the profile of a Gaussian beam, taking into account a Gaussian atom distribution with $\sigma_c = \langle \rho^2 \rangle \simeq 0.6$ mm. This results in

$$\frac{\delta k}{k} = \frac{r_I^{(2)} w_0^2}{16k^2\sigma_c^2}\left(\frac{I_c}{I_0}\right)^{w_0^2/4\sigma_c^2-1},$$

where $r_I^{(2)}=\nabla_\perp^2\, r_I(0)$ is the Laplacian of the 2-D autocorrelation function $r_I(\mathbf{x})$ of the intensity fluctuations, and $I_0$ the beam intensity at the center.

We calculate the autocorrelation function $r_{\mathrm{I}}(\mathbf{x})$ from CCD intensity profiles taken at the approximate distance of the atoms from the fiber port where they interact with the laser beam (Fig. S1). The rms amplitude of the noise is given by the autocorrelation function at the origin, $r_{\mathrm{I}}(0)^{1/2} \simeq 0.65\%$ of the beam intensity, and the Laplacian of the autocorrelation function at the origin is measured to be $r_{\mathrm{I}}^{(2)} \simeq 2.44 \times 10^{-9}\ \mu\mathrm{m}^{-2}$. With $I_c = 0.85 \pm 0.05\ I_0$, we find the magnitude of this systematic effect to be -0.030 ± 0.019 ppb.

We can also check the validity of this result by directly measuring the shift in α as the Bloch intensity is varied. The measured dependence on the Bloch efficiency is shown in Figure S5. No dependence is observed within 0.3 ppb (limited by statistics), consistent with the results of the analytic calculation and Monte Carlo.

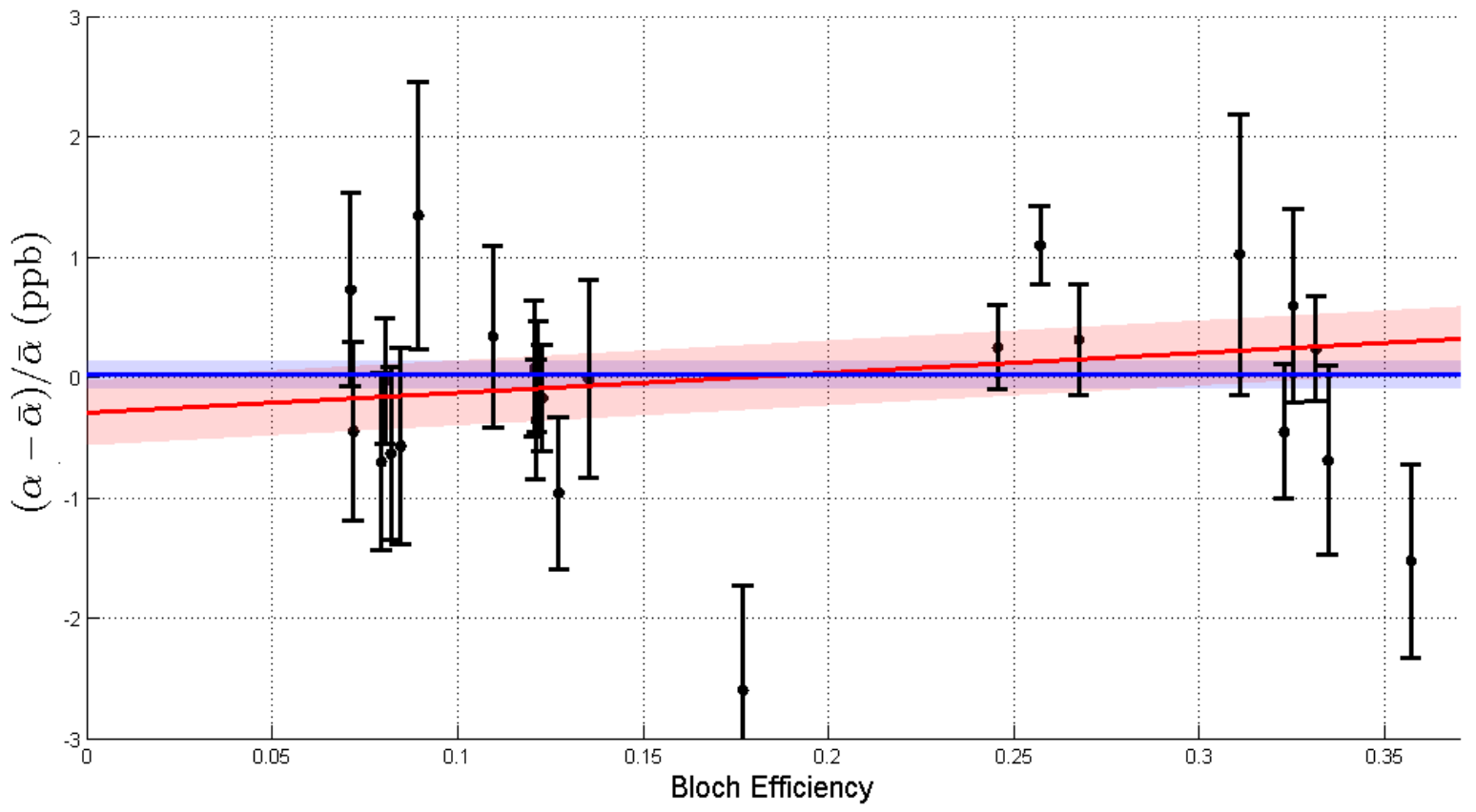

**Fig. S5. Varying Bloch Intensity.** Experimental measurements of α, as the Bloch efficiency (i.e. intensity) is varied. The blue line is the final reported value of α, assuming no dependence on Bloch efficiency. The red line assumes a linear dependence; the shaded regions represent 1-sigma error bars.

As a final check that the Gouy phase was accounted for correctly, data was taken with two different beam waists (the 5.7 mm beam and the apodized beam). The α values from these two measurements were within 1-sigma (Figure S6).

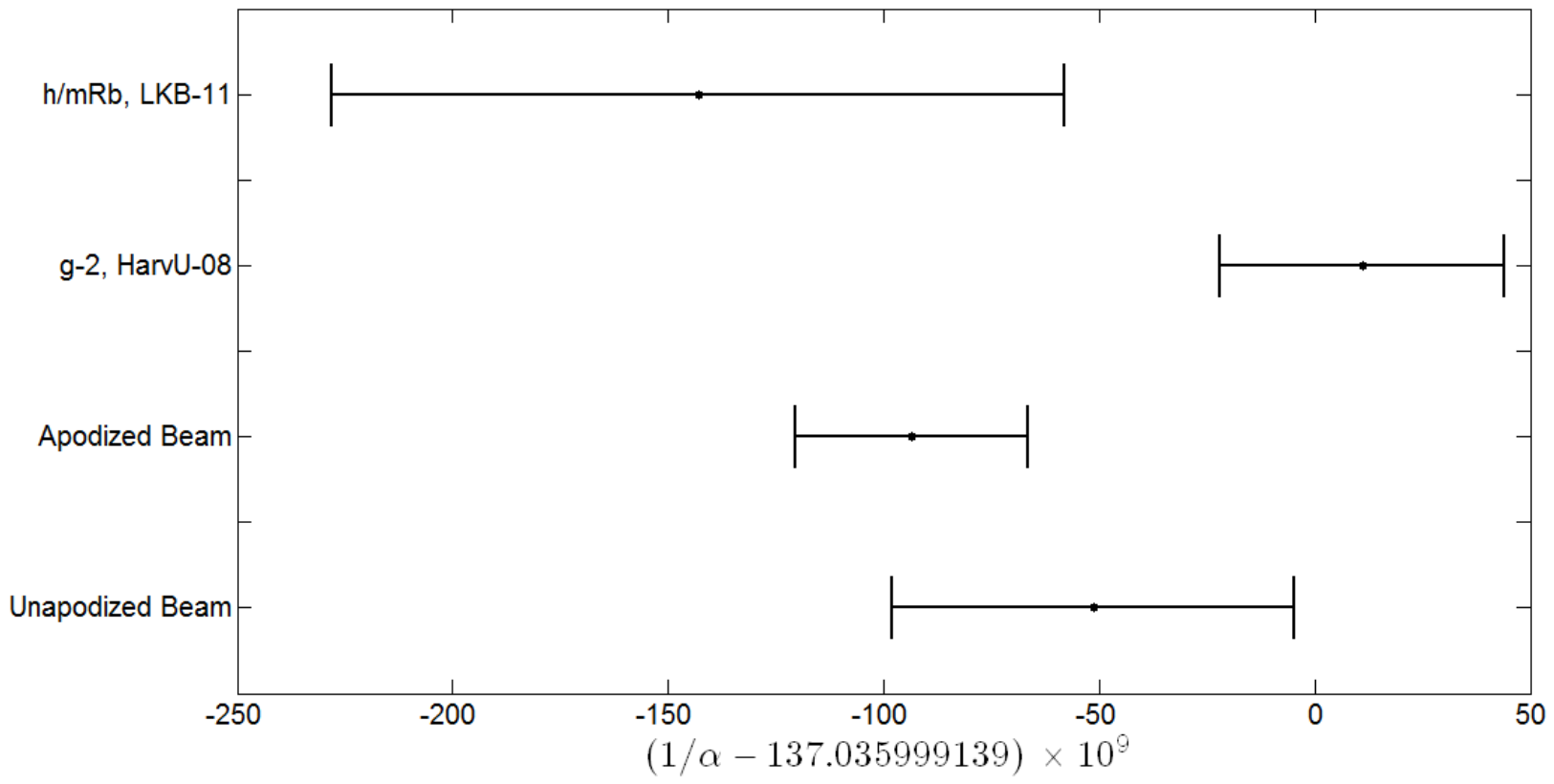

**Fig. S6. Comparing Beam Waists.** Experimental measurements of α, for two different beam sizes (the unapodized 5.7 mm beam, and the apodized flat-top beam). Also plotted are the LKB-11 and $g_e$-2 measurements of α for reference.

## Section 4: Spatially dependent potentials

Spatially varying potentials that act on the atoms may cause systematic or stochastic fluctuations of the phase Φ which depend on the atoms' position and thus on time as the atoms move in the interferometer. Systematic contributions may arise, e.g., from gravity and magnetic fields. Stochastic fluctuations result, e.g., from speckle, random variations of the laser intensity and phase.

### A: Acceleration gradient from gravity and magnetic fields

While the simultaneous conjugate Ramsey-Bordé interferometer cancels the effects of a constant acceleration, the fractional error in the recoil frequency due to the gravity gradient $\gamma=\partial g_z/\partial z$, where $z$ is the vertical coordinate, is

$$\Phi_{\text{gradient}} = \frac{4}{3}\gamma n \omega_r T \cdot \left[ n\left(2T^2 + 3TT_2' + 3\left(T_1'^2 + T_2'^2\right)\right) + 2N\left( T^2 + 3T\left( T_2' - \frac{NT_\text{B}}{2}\right) + 3T_2'\left(T_2' - NT_\text{B}\right) + \left(N^2 - \frac{1}{4}\right)T_\text{B}^2 \right)\right] .$$

We measure $\gamma$ using a gradiometer consisting of two vertically separated Mach-Zehnder interferometers (*32*), as shown in Fig. S7. The differential phase of this configuration is

$$\Delta\Phi_\gamma = 8n\omega_r\gamma T^2\left[N\left(2T - NT_B + 2T_2'\right) + n\left(T + T_1' + T_2'\right)\right] + \frac{8gn(n+2N)T^2\omega_r}{c},$$

where $T_B$ is the duration of a single Bloch oscillation. We take data at $N$=125, $T'_1$+$T'_2$=50ms, with pulse separation times $T$ varying between 60 and 100ms.

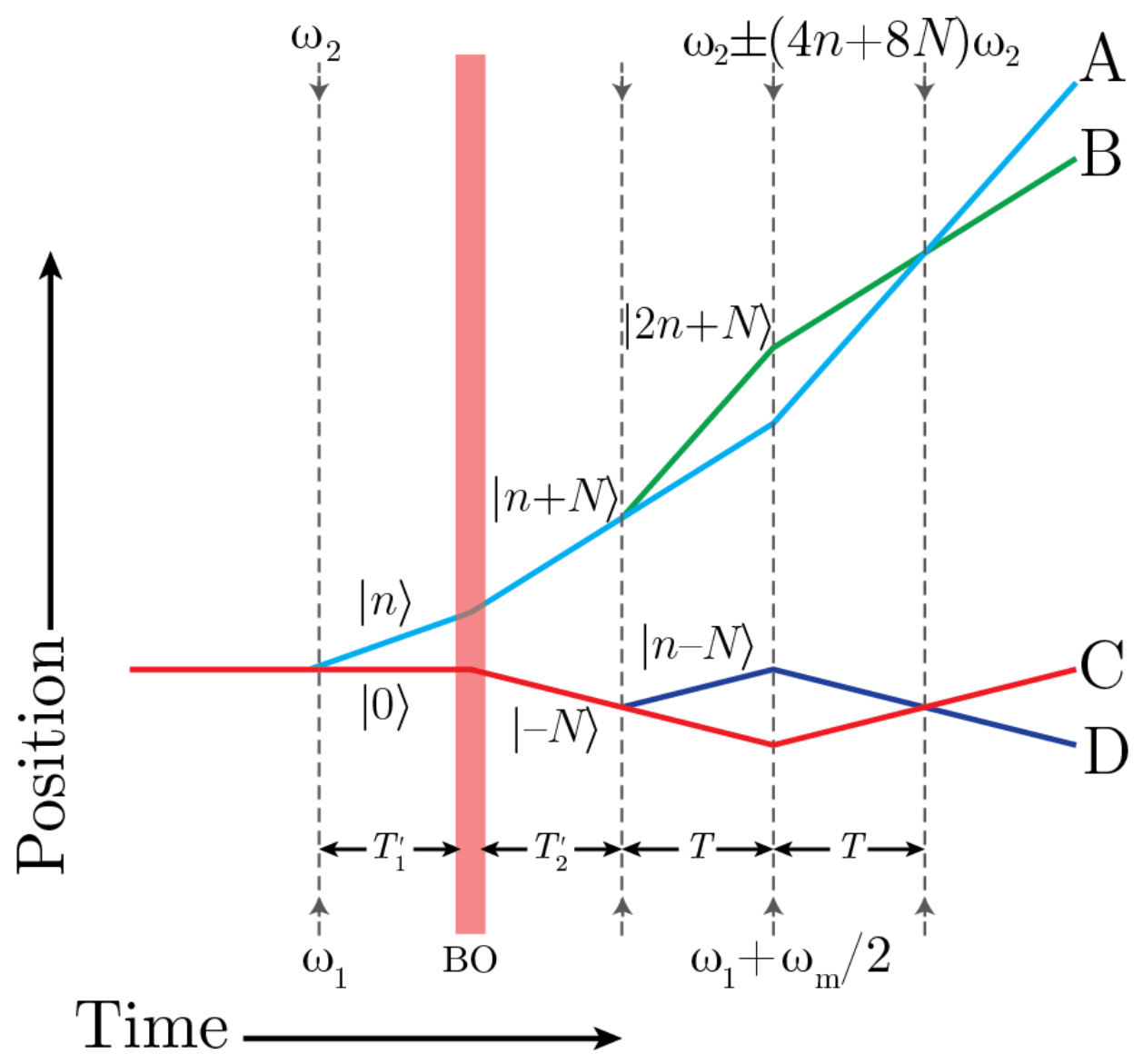

**Fig. S7. Gradiometer Geometry.** The interferometer geometry used to measure the gravity gradient.

We also consider the effect of second-order variations of gravity, the gradient of the gradient. This will be dominated by the local mass distribution, particularly the $M \sim 15$ kg detection chamber below the atom interferometer. As the atoms are never closer than $r$ = 40 cm from the detection chamber, we can calculate the gradient to be at most $2GM/r^3 = 2.9 \times 10^{-8}\ s^{-2}$ at the closest approach of the atoms, decaying rapidly with distance. If the atom interferometer measuring α and the gravity gradiometer are sensitive to the gradient at the same effective location, this cancels out between the two measurements, but these locations differ by 5 cm. (Additional suppression is provided by the fact that this extra gradient drops sharply with distance; we will not consider this.) This results in a contribution of 0.01 ppb, which is added in quadrature to the error from the gradiometer measurement for an overall uncertainty of 0.02 ppb. Objects further away have even smaller influences that we neglect. For example, 1000 kg at $r$ = 2 m (the optical table weighs about 700 kg) lead to 0.003 ppb and $6\times10^3$ kg (an estimate for the weight of $(2.5\ m)^2$ of the floor) at 2.5 m to 0.002 ppb.

The effects of magnetic fields can be fully accounted for as a contribution to the acceleration gradient. Our atoms are in the $F$=3, $m_F$ = 0 state and only experience a quadratic Zeeman shift of about β=+213 Hz/G$^2$. The magnetic field in the interferometer region can be modeled as a polynomial $B(z)=B_0+B'z+B''z^2+\dots$ The corresponding energy shift changes the Lagrangian for the atoms by:

$$L_B = \frac{g_J^{\,2}\mu_B^{\,2}}{4\Delta E_{\mathrm{hfs}}} B^2(z) \approx \hbar\beta\left[B_0^{\,2} + 2B_0B'z + \left(B'^2 + 2B_0B''\right)z^2\right].$$

The $B_0^2$ term is common mode to all arms of the interferometer and can be ignored. Comparing this Lagrangian to the one due to gravity, $L = p^2/(2m)-mgz+m\gamma z^2/2$, shows that the term linear in $z$ is similar to the one caused by a linear gravitational potential and cancels out between the two interferometers. The term proportional to $z^2$ causes an acceleration gradient and can be absorbed into the gravity gradient term by substituting $\gamma \rightarrow \gamma+2\beta(B'^2+2B_0B')/m$. By applying gravity gradient corrections from the gradiometer measurements, we have already dealt with these magnetic gradients.

As an independent verification of this approach, data on the fine structure constant was taken with bias B-fields of 0.38 G and 3.7 G, with the resulting recoil frequencies consistent with each other to within 1-sigma (1.4 ppb). This puts an upper-bound on any systematic due to magnetic fields at the smaller bias field at 0.014 ppb, which is further reduced after the acceleration gradient is measured and taken out.

B: Speckle

After the corrections for the acceleration gradient and other systematics have been applied, the measurement of α should be independent of the pulse separation time $T$. In the initial phase of the experiment, however, we observed anomalous variations of the interferometer phase Φ that could be as large as 30 mrad and that varied between fountain re-alignments (Fig. S8, red graph).

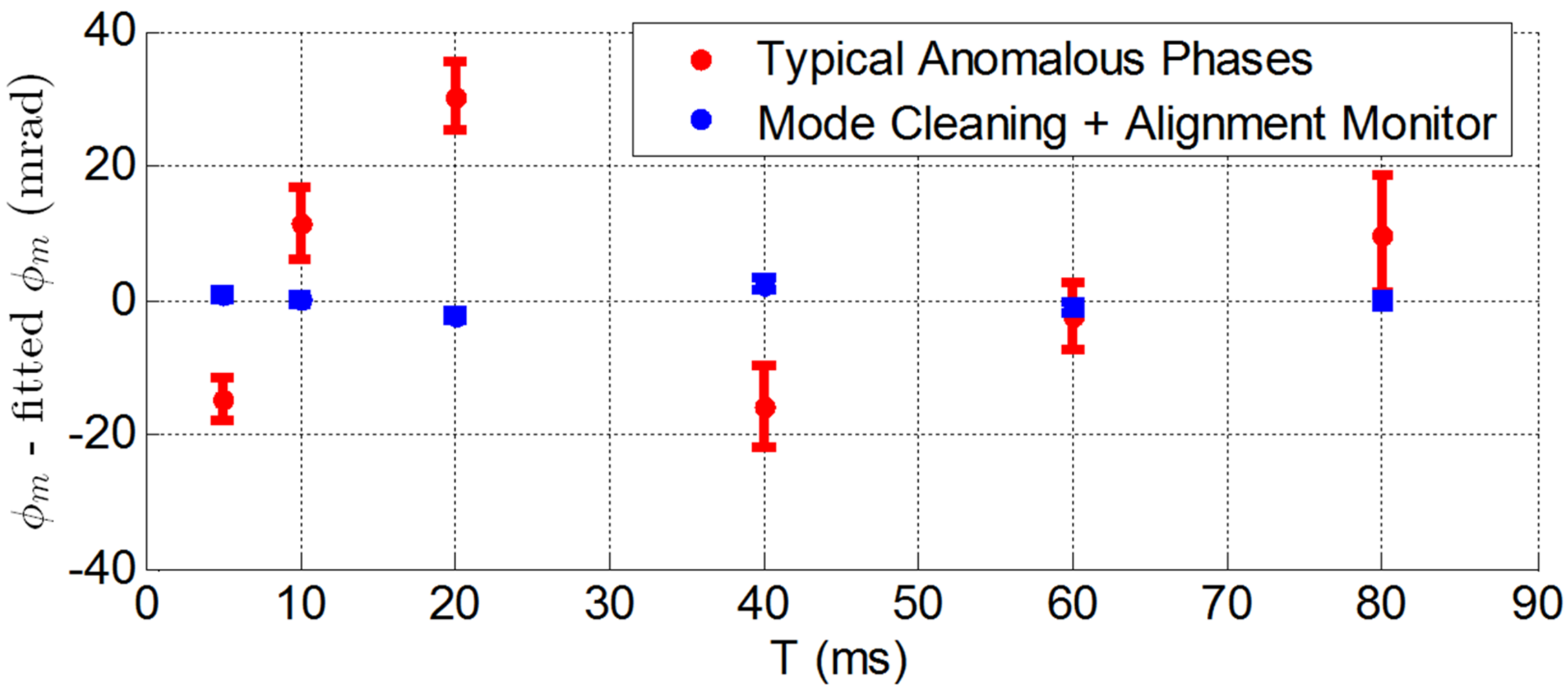

**Fig. S8. Speckle Phase Suppression.** Anomalous phases measured after subtracting off all other known systematic errors, with $N = 0$. With no effort taken to suppress them, the speckle phases can be as large as 30 mrad (red data). With the real-time fountain monitor and apodizing filter, the speckle phases are kept below 3 mrad.

We traced the anomaly to random intensity variations of the laser beam used for Bragg diffraction and Bloch oscillations. This beam enters the vacuum system after exiting a single-mode, step-index optical fiber, which delivers a Gaussian intensity profile near the beam axis, out to about one Gaussian waist where the intensity reaches 10% of the peak. Beyond that, the beam is approximately Lorentzian-shaped and thus has much higher intensity far away from the center than a Gaussian. This intensity reflects off the vacuum chamber walls and interferes with the main beam, causing irregular spatial variations of the beam intensity.

Reducing the amount of scattered light was found to suppress the anomalous phases. To this end, we added an apodizing filter (Thorlabs NDYR20A) to the output of the fiber port, gradually attenuating the intensity as a Gaussian function of distance from the beam axis. The effectiveness of this approach is shown in Figure S6 (blue graph): the anomalous phase can be kept below 3 mrad.

This strategy will not suppress light scattered from dust or scratches, however, so small anomalous phases may still be present. Empirically, we found the speckle phase shifts to be independent of the Bloch order; therefore we can reduce their fractional significance by increasing the Bloch order. For example, at $N$=50 the anomalous phases need to kept below 4 mrad for an 0.25 ppb measurement of α—for this reason, only data with $N$=125 or larger is used in the determination of α.

C: Conclusion and verification

Anomalous phase variations with $N = 0$ and without mode cleaning (Fig. S7, red) would cause an about ± 8 ppb contribution to the experimental uncertainty. With mode cleaning (Fig. S7, blue), the effect is reduced by an order of magnitude. Operating at $N = 125$ reduces the effect by a factor $n/(n+N) = 0.04$, and therefore we expect a possible contribution of the anomalous phase to the error budget of approximately 0.03 ppb. In order to verify this residual effect of any (unresolved) anomalous phases, residual stochastic variation in the data, we implement a model to estimate the error under the assumption that the residuals are entirely due to speckle phases (and not due to random statistical fluctuations). The systematic error for the residuals shown in Figure S9 is below 0.04 ppb in α.

The data in Figure S9 verifies both our understanding of systematic shifts and the speckle phase. Any anomalous phase shifts that depend on the pulse separation times $T$ are suppressed to the point where any remaining anomalous phases are unresolved.

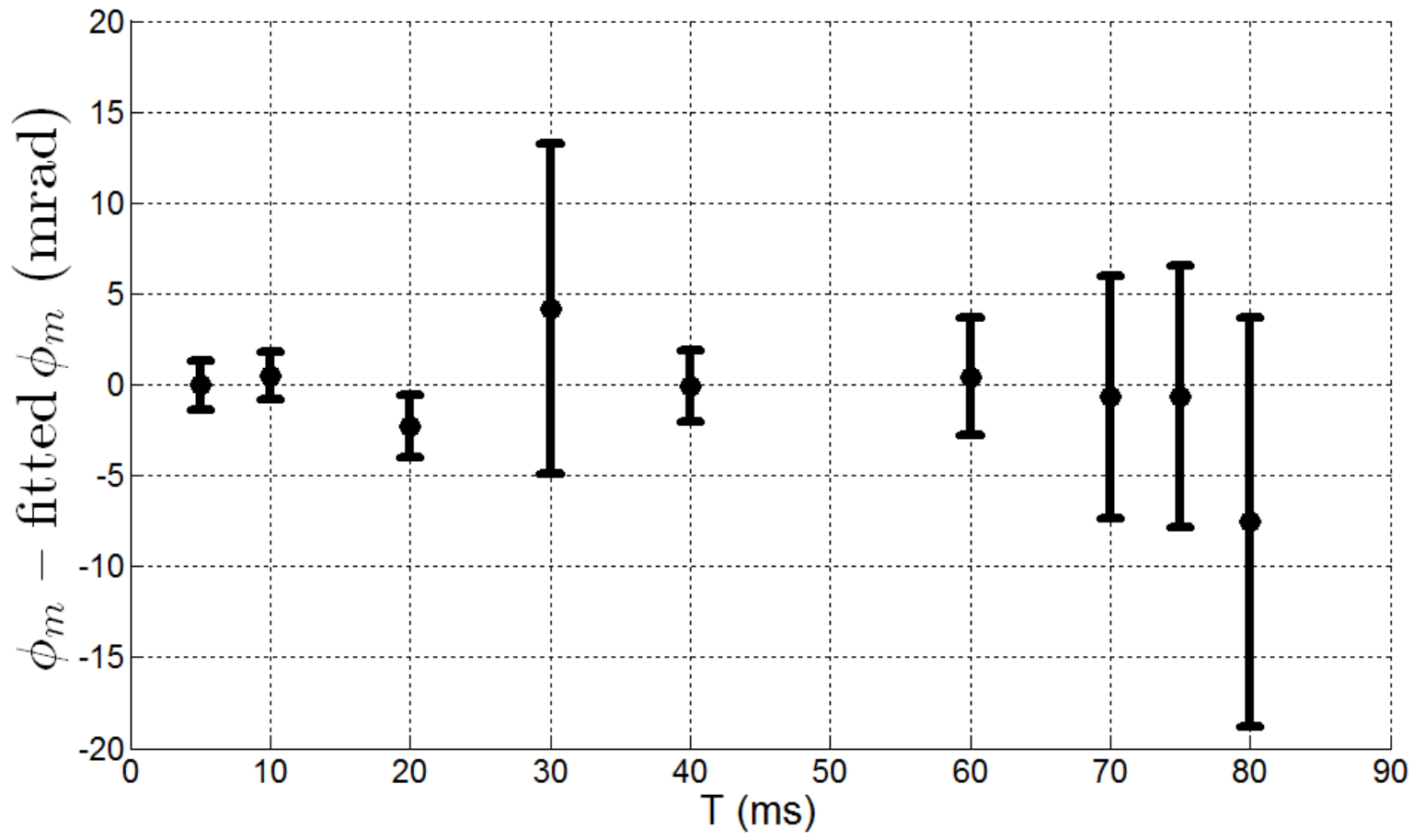

**Fig. S9. Speckle Phase Suppression.** Anomalous phases measured after subtracting off all other known systematic errors, with the real-time fountain alignment monitor, apodizing filter, and $N \geq 125$; these are the residuals vs $T$ for the data shown in Fig 3C.

Section 5: Beam Alignment Correction

Misalignment of the retro-reflection angle θ of the Bragg beam reduces the effective wave vector as ${k_{\mathrm{eff}}}^2 \approx 2k^2(1+\cos\theta) \approx 4k^2 - k^2\theta^2$. To measure θ, we monitor the back-coupling efficiency of the light to the fiber. We calibrated the back-coupling efficiency as a function of angle. By measuring the coupling efficiency during the experimental run, the relative angle (typically around 12 μrad) can thus be determined which allows for post correction of the data to reduce the beam alignment systematic.

Section 6: Bloch Oscillation Light Shift

During Bloch oscillations, the atoms are shifted in energy by an amount

$$\delta E_{ac} = \frac{\hbar\Omega^2}{4\Delta}$$

due to the ac-Stark effect, where Ω is the (local) Rabi frequency of the optical lattice on the atoms and Δ is the single-photon detuning. This light shift enters the phase of the interferometer. The energy shift applies for each component of the Bloch lattice beam for a total of 6 beams, as the three frequencies $\omega_1$, $\omega_\pm$ travel up and down after retro-reflection. Since the Bloch lattice is blue-detuned, the atoms sit at a potential minimum of the lattice used to accelerate them. The dominant ac Stark shift is thus caused by the time-averaged energy shift from the remaining four beams, which is a function of the beam intensities at the location of the atoms.

Denote $I_1^\uparrow(z)$, $I_\pm^\uparrow(z)$ the intensities of these frequencies as function of $z$ in the upgoing laser beam and $I_1^\downarrow(z)$, $I_\pm^\downarrow(z)$ the same in the downgoing beam. Denote $z_{1\text{-}4}$ the positions of the partial wave packets during Bloch oscillations, measured relative to the top mirror, where $z_1$ is the position of the initially uppermost wave packet and $z_4$ the initially lowest.

For the atoms being accelerated upwards, the dominant ac-Stark shift arises from $I_\pm^\uparrow$ and $I_1^\downarrow$, $I_+^\downarrow$. For the atoms being accelerated downwards, the dominant ac-Stark shift arises from $I_\pm^\uparrow$ and $I_1^\downarrow$, $I_-^\downarrow$. Defining the abbreviation $J = I_-^\uparrow + I_+^\uparrow + I_1^\downarrow$, the total ac Stark shift entering the interferometer phase is proportional to $\Delta I = J(z_1) - J(z_3) - J(z_2) + J(z_4) + I_+^\downarrow(z_1) - I_+^\downarrow(z_3) - I_-^\downarrow(z_2) + I_-^\downarrow(z_4)$. This is valid for any shape or intensity profile of the laser beam.

If there are stochastic variations of intensity with $z$, we will observe stochastic variations of the interferometer phase with $T$. Such variations are bounded experimentally and their effect on our result is described in section 4B and 4C.

A systematic shift can arise from the propagation of our Gaussian laser beam. To determine it, we write $z_1=z_4+d+\delta$, $z_2=z_4+d$, $z_3=z_4+\delta$, where $d = 2\, n\, v_r\, T$ is the splitting of the interferometer arms and δ is the distance between the two interferometers during Bloch oscillations. We expand the intensities as $I(z_4+\epsilon)=I+I'\epsilon+I''\epsilon^2/2$, where $I, I', I''$ are the intensities and their derivatives taken at $z_4$. This yields

$$\Delta I \approx \left[\left(I_+^{\downarrow}\right)' - \left(I_-^{\downarrow}\right)'\right] d + \left[J'' + \left(I_+^{\downarrow}\right)''\right] \delta d + \frac{1}{2}\left[\left(I_+^{\downarrow}\right)'' - \left(I_-^{\downarrow}\right)''\right] d^2.$$

The dominant contribution is the first term. The two beams involved come from the same optical fiber and were both retroreflected at the mirror. The $z$-dependence of their intensities as well as any retro-reflection losses are thus common to both, and the derivatives are

$$\left(I_\pm^{\downarrow}\right)' = I_{0,\pm} \frac{2}{z_R^2} \frac{z_4 - z_0}{\left(1 + \frac{\left(z_0 - z_4\right)^2}{z_R^2}\right)^2},$$

where $z_0$ and $z_R$ are the common location of the waist and the Rayleigh range of these beams and $I_{0,\pm}$ are the intensities of the two beams at the waist. With $z_R$=35.0 m, $z_0$-$z_4$=1.5 m, $n$=5, $v_r$=3.5 mm/s, $T$=80 ms, we obtain $\Delta I \approx 6.8 \times 10^{-6}(I_{0,+}-I_{0,-})$.

Each beam by itself causes an AC Stark shift of about $1.5\hbar\omega_r$ and within the 8-ms duration of our Bloch oscillations causes a phase shift of about 0.16 krad. With a 2% intensity balance between $I_{0,+}$ and $I_{0,-}$ (verified by comparing the Bragg diffraction efficiency when kicking atoms up and down), we estimate $\Delta_{ac}\Phi$= 0.16 krad × 6.8 × $10^{-6}$ × 0.02 ~ 22 µrad, a 0.002 ppb change in α for $n$ = 5, $N$ = 125, $T$ = 80 ms.

Section 7: Density Shift Correction

The energy shift in our cloud due to density shifts can then be approximated as $E_\rho = \rho 4\pi\hbar^2 a_s / m$, where $a_s$ = 280(10)·$a_0$ for cesium in the $F$=3 state and $a_0$ is the Bohr radius *(30)*. For typical interferometer signals, the atom number density during detection is approximately $2.5\times10^5$ atoms/cm$^3$, and the density during the interferometer sequence is approximately $10^6$ atoms/cm$^3$. For an exaggerated beam splitter imbalance of 3:1, the differential density shift will cause a net interferometer phase of

$$\Delta\Phi = \frac{\Delta E_\rho}{\hbar}\left(2T + T'\right) = \left(\frac{3\rho}{4} - \frac{\rho}{4}\right)\frac{4\pi\hbar a_s}{m}\left(2T + T'\right)$$

For typical parameters of $T = 80$ ms and $T' = 10$ ms, this gives a phase of 8 μrad corresponding to a negligible 0.003 ppb shift in $\omega_r$.

Section 8: Index of Refraction Correction

The effective wavenumber $k_{\mathrm{eff}}$ of the laser is sensitive to the refractive index in the vacuum chamber, arising from room-temperature as well as cold background cesium atoms. The refractive index in an atomic vapor in the far-detuned limit is

$$n - 1 = \frac{\sigma_0 \rho \Gamma}{4k\Delta} \ ,$$

where $\sigma_0 = 2.5 \times 10^{-9}$ cm$^2$ is the resonant cross section, $\rho$ is the atom number density, $\Gamma$ is the natural linewidth, $\Delta \approx 14$ GHz is the single photon detuning, and $s$ is the saturation parameter $I/I_{\mathrm{sat}}$. As an upper bound for the density of cold cesium atoms, we use the density of the atom cloud during the interferometer sequence, $10^6$ atoms/cm$^3$; the corresponding index of refraction is below $n$-1 = 0.003 ppb, which we take to be negligible. The dominant contribution comes from room-temperature background atoms. An upper bound of $10^7$ atoms/cm$^3$ on their density is obtained by assuming that the entire vacuum pressure of $10^{-9}$ torr is due to cesium, and the corresponding index of refraction shift is at most $n$-1 = 0.03 ppb. For our low atom densities, no detailed simulation of the atom cloud is needed.

Section 9: Sagnac Phase Systematic

Our Ramsey-Bordé interferometer configuration ideally has zero enclosed spatial because all motion takes place in the vertical direction. If there is a misalignment such that the two interferometer paths enclose a spatial area **A**, then there will be an extra phase shift

$$\Phi_\Omega = \frac{4\pi m}{\hbar}\vec{A}\cdot\vec{\Omega}$$

due to the Sagnac effect, where **Ω** is the rotation vector. The enclosed areas of the upper and lower interferometers mostly cancel, but a difference arises because of the gravity gradient γ if the laser beams are rotating in the atom's inertial frame at a rate ω as a result of imperfect Coriolis compensation. The laser rotation rate is zero with perfect Coriolis compensation and ω

= $\Omega_e$ (i.e. the rotation rate of the Earth) without compensation. The difference in area between the upper and lower interferometers is then

$$A_u - A_l = -n(n+N)v_r^{\,2}T^2\left[\frac{4}{3}\left(12+\gamma T^2\right)-(12+5\gamma T^2)\cos\omega T\right]\sin\omega T\,,$$

which is below 0.001 ppb after being cancelled to below 10% using Coriolis compensation.

Section 10: Modulation Frequency Wavenumber

During the third and fourth beam splitters of a simultaneous conjugate Ramsey-Bordé interferometer, the effective wave-vector for the upper and lower interferometers differ by $\pm\omega_m/c$ due to the extra modulation frequency added to the laser to drive both Bragg orders. This perturbation in the wave results in additional phases

$$\Phi_{\delta k} = \frac{gnT(3T+2T_1'+2T_2')\omega_m}{c} - 4n^2\omega_r T\frac{\omega_m}{\omega_L}$$

for $\omega_m \approx 8(n+N)\omega_r$, For typical parameters $T$ = 80 ms and $T'$ = 10 ms, the correction is -4.25 ppb. Since the value of $g = \vec{g}\cdot\vec{k}\,/\,|k|$ is the projection of gravity onto the wave-vector, the vertical alignment of the Bragg beam introduces a slight uncertainty, but even with a misalignment as large as 10 mrad, it is below 0.001 ppb.

Section 11: Systematics Due To The Atoms' Thermal Motion

As described in (*17*), the differential diffraction phases for different output ports of the simultaneous conjugate interferometer do not produce a systematic effect to leading order. However, there are higher-order effects that can produce a systematic effect by causing the diffraction phase to vary with the pulse separation time *T*. The dominant effect comes from the thermal motion of the atoms as they a ballistically expand—this causes the atoms to see a different Bragg pulse intensity at different times during the interferometer sequence. Expanding the diffraction phase $\Phi_0$ to first order in *T* results in an extra phase term when the total interferometer phase $\Delta\Phi = 0$, so that now the interferometer output is given by

$$2n\omega_m T = 16n(n+N)\omega_r T + \Phi_0 + \frac{d\Phi_0}{dT}T\,.$$

To calculate $d\Phi_0/dT$, a Monte Carlo simulation is used, which is based on a Gaussian density and velocity profile of the initial atomic cloud. The simulation takes into account the spatial filtering pulses as well as the Bragg pulses at the location of each atom and is described in detail in (*17*). The simulation is run for the particular experimental parameters (Bragg pulse intensity, cloud temperature, etc.) used in this work, as well as 1-sigma variations in those parameters limited by the experimental repeatability. The resulting systematic shifts are presented in Table S1; the simulation is run enough times so that each shift is well-resolved compared to the numerical error bar. In total the thermal motion of the atoms introduces a systematic uncertainty of 0.08 ppb in α.

**Table S1.** Systematic Shifts Due to Ellipse Distortion. The table shows the results of a Monte Carlo simulation quantifying the systematic shifts arising from thermal motion of the atoms, which introduces distortion in the ellipses used for phase extraction. The parameters used in the model are allowed to vary, replicating the level of control achieved in the actual experiment.

| Effect | Value | $\delta\alpha/\alpha$ (ppb) |
|---|---|---|
| Cloud radius (mm) | 2.2 ± 1 | ± 0.026 |
| Vertical velocity width (vr) | 1.5 ± 0.25 | ± 0.031 |
| Ensemble horizontal velocity (vr) | 0 ± 0.5 | ± 0.032 |
| Initial horizontal position (mm) | 0 ± 1 | ± 0.034 |
| Intensity ($I_{\pi/2}$) | 1.02 ± 0.02 | ± 0.028 |
| Last pulse intensity ratio | 1.0 ± 0.02 | ± 0.034 |

Note that the 'cloud radius' in this table refers to the density distribution alone, not the contrast-weighted density distribution used to determine the effective Gouy phase (Section 3).

Section 13: Non-Gaussian Waveform

The *T*-dependent diffraction phases described in the previous sections can be amplified by imperfections in the experimental setup, particularly if the temporal waveform used for the Bragg diffraction is significantly non-Gaussian. A detailed treatment is given in (*17*); by using an intensity servo to stabilize the temporal waveform to a reference Gaussian waveform, this systematic effect can be kept below 0.03 ppb.

Section 14: Parasitic Interferometers

As discussed in (*17*), the multi-port nature of Bragg diffraction allows for the formation of unwanted Ramsey-Bordé interferometers that will close at the same time as the main interferometer and will not be suppressed by Bloch oscillations. These unwanted interferometers will produce small, oscillating phase shifts as the pulse separation time $T$ is varied, and can produce a systematic shift as large as 1 ppb in α. Using the Monte Carlo described above, it was determined that the dominant contribution to this effect comes from unwanted population in the $n$=1 order, which can be suppressed by appropriate choice of the Bragg pulse duration (109 μs for our experimental parameters). The simulation's prediction of the parasitic phases is in good agreement with experimental data, and the expected systematic shift at the 'magic' Bragg duration is below 0.03 ppb in α. See Ref. (*17*) for more detail.

Section 15: Doppler Correction

The frequency difference $\omega_m$ between the frequency pairs sent towards the atoms is seen Doppler-shifted by the free-falling atoms. In the lab frame, this causes an additional Doppler phase of $\Phi_{\mathrm{Doppler}} = -2nT\omega_m v_0/c$, so that $\tilde{\omega}_m = \omega_m\left(1-v_0/c\right)$. Here $v_0$ is the atom velocity at the first beam splitter.

Effects like the frequency modulation of the 3rd and 4th pulses and the gravity gradient will cause the atom interferometer to not fully close. To account for this in the phase calculation, in addition to the free-evolution phase and laser-interaction phase, a splitting phase is added either at the beginning or end of the interferometer, given by $\Phi_{\mathrm{Splitting}} = \mathbf{p}_0 \cdot \Delta\mathbf{z}/\hbar$, where $\mathbf{p}_0$ is the atom momentum and $\Delta\mathbf{z}$ is the atom path separation. $\mathbf{p}_0$ is determined by the particular reference frame chosen, and therefore is not defined at a particular pulse; when the frame transformations are accounted for, $\Phi_{\mathrm{Splitting}}$ will cancel in the interferometer sequence and therefore does not appear in Eqn. S1.

These two velocity-dependent phase terms do not violate Lorentz symmetry: In the atom frame, the splitting is zero to leading order, but the modulation frequency seen by the atoms is different from the one measured in the lab, while in the lab frame, the splitting is nonzero. The total correction $\Phi_{Doppler} + \Phi_{Splitting}$ is frame-invariant, and results in a correction $v_0/c$~7 ppb to α (Section 1). Because $v_0$ can be measured by adjusting the laser frequency to find Bragg resonance, it is known to better than 1 part per thousand and does not introduce additional systematic uncertainty.

Section 16: $g_e$-2 Corrections and Dark Matter Limits

The experiment in this work determines α by measuring the ratio $h/m$. Another method involves measuring the electron gyromagnetic anomaly $g_e$-2, which can be written as a power series in α using corrections from quantum electrodynamics (QED) as:

$$g_e - 2 = \sum_{n-1} \left( \frac{\alpha}{\pi} \right)^n a_n + \alpha_{weak} + \alpha_{QCD}$$

where the coefficients $a_n$ come from calculating all possible QED corrections of order $2n$ and the factors $\alpha_{weak}$ and QCD are electroweak and quantum chromodynamics corrections obtained from particle physics data (*4*). Thus a comparison of the two kinds of experiments can be used as a test of the standard model of particle physics. Comparison of our experimental result with the most precise value of $g_e$-2 obtained through direct measurement (*2*) yields a negative $\delta a = a_{meas} - a(\alpha) = -0.88(0.36) \times 10^{-12}$. The relative magnitude of the different corrections that go into calculating $g_e$-2 are shown in Figure S10; the two experiments have an error bar below the magnitude of the 5th order QED correction, however the sign of the experimental discrepancy is opposite that of the 5th order correction.

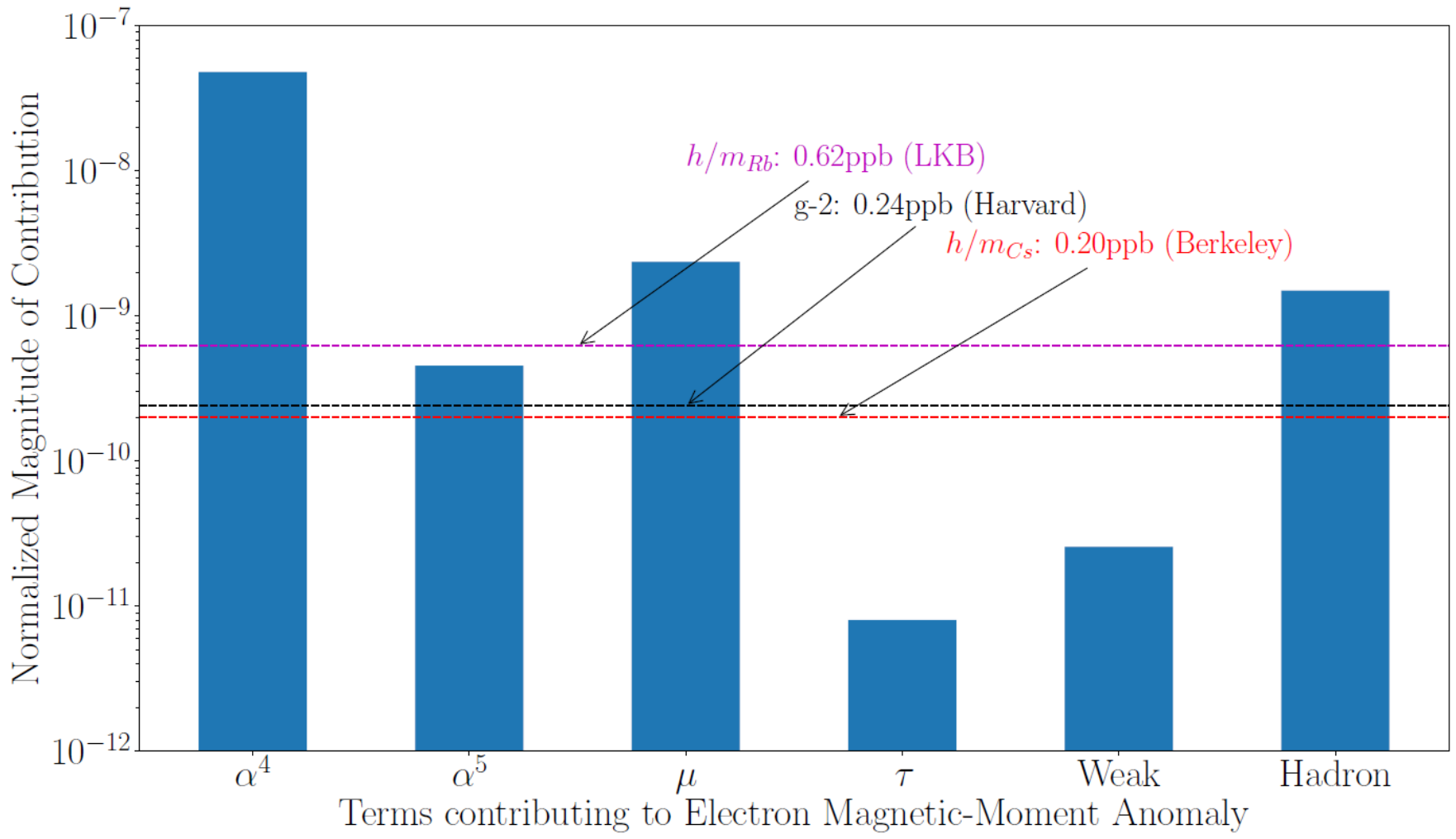

**Fig. S10. $g_e$-2 corrections.** The relative magnitude of the different corrections that go into calculating the electron anomalous magnetic moment $g_e$-2, with dashed lines showing the experimental precision achieved by the LKB (*4*) and Harvard (*2*) experiments, as well as this work (labeled as 'Berkeley'). For clarity, the much-larger $\alpha^1$, $\alpha^2$, and $\alpha^3$ terms are not shown.

Precision measurements of α like ours can also help searching for new dark-sector particles; in particular we consider dark vector and axial-vector bosons. Following the treatment of (*28*), we consider the general Lagrangian for a massive gauge boson $A'$ with both vector and axial vector couplings:

$$\mathrm{L}_{A'} = -\frac{1}{4}F'_{\mu\nu}F'^{\mu\nu} - \frac{m_{A'}^2}{2}A'_{\mu}A'^{\mu} + A'_{\mu}\sum_f \bar{f}\left(c_V^f\gamma^\mu + c_A^f\gamma^\mu\gamma^5\right)f,$$

where $F'_{\mu\nu}=\partial_\mu A'_\nu-\partial_\nu A'_\mu$ is the field strength tensor, $f$ is an SM fermion and $c^f_{V,A}$ are its vector and axial-vector couplings respectively. The vector electron coupling of dark photons contributes to the anomalous magnetic moment of the electron as

$$\delta a_e = \frac{\left(c_V\right)^2}{4\pi^2}\int_0^1 \frac{x^2(1-x)}{x^2+\frac{m_V^2}{m_e^2}(1-x)}dx$$

where the coefficient $c_V^2/4\pi^2$ is sometimes written as $\alpha\epsilon/2\pi$, where $\epsilon$ is the mixing strength. And the axial-vector coupling contributes as

$$\delta a_e = \frac{\left(c_A\right)^2}{4\pi^2}\frac{m_e^2}{m_{A'}^2}\int_0^1 \frac{2x^3+\left(x-x^2\right)\left(4-x\right)\frac{m_{A'}^2}{m_e^2}}{x^2+\frac{m_{A'}^2}{m_e^2}(1-x)}dx.$$

We generate the exclusion plots in Figure 4 by requiring either the pair ($c_V$, $m_V$) or ($c_A$, $m_A$) are consistent with the measured $\delta a$; the vector limits are calculated assuming no axial-vector coupling, and vice-versa.